\documentclass[a4paper]{article}
\usepackage[english]{babel}
\usepackage[utf8]{inputenc}
\usepackage{amsmath, amssymb, amsthm}
\usepackage{graphics}
\usepackage{todonotes}
\usepackage{physics}
\usepackage[margin=0.9in]{geometry}
\usepackage{hyperref}

\usepackage{floatrow}

\usepackage{caption}
\usepackage{subcaption}
\usepackage{floatrow}
\usepackage{xpatch}

\usepackage{graphicx}
\usepackage{mathrsfs}
\usepackage{verbatim}

\newcommand{\Z}{\mathbb{Z}}
\newcommand{\arctanh}{\text{arctanh}}
\newcommand{\cc}{\leftrightarrow}

\usepackage{bbm}
\DeclareMathOperator{\id}{\mathbbm{1}} 
\DeclareMathOperator{\win}{in} 

\newtheorem{theorem}{Theorem}[section]
\newtheorem{definition}[theorem]{Definition}
\newtheorem{lemma}[theorem]{Lemma}
\newtheorem{remark}[theorem]{Remark}
\newtheorem{corollary}[theorem]{Corollary}
\newtheorem{conjecture}[theorem]{Conjecture}

\newtheorem{proposition}[theorem]{Proposition}

\newlength\tindent
\title{Strict monotonicity, continuity and bounds on the Kert{\'{e}}sz line for the random-cluster model on $\Z^d$}
\author{Ulrik Thinggaard Hansen \\ 
Department of Mathematics \\
Université de Fribourg, Chem. de Musée 9, 1700 Fribourg, Switzerland \\
\\
Frederik Ravn Klausen\footnote{ klausen@math.ku.dk} \\
QMATH, Department of Mathematical Sciences,
\\ University of Copenhagen, Universitetsparken 5,DK-2100
Copenhagen Ø, Denmark
 }
\date{\today}

\begin{document}

\maketitle 
\begin{abstract}

Ising and Potts models can be studied using the Fortuin–Kasteleyn representation through the Edwards-Sokal coupling. This adapts to the setting where the models are exposed to an external field of strength $h>0$. In this representation, which is also known as the random-cluster model, the Kert{\'{e}}sz line is the curve which separates
two regions of the parameter space defined according to the existence of an infinite cluster in $\mathbb{Z}^d$. This signifies a geometric phase transition between the ordered and disordered phases even in cases where a thermodynamic phase transition does not occur. 
In this article, we prove strict monotonicity and continuity of the Kert{\'{e}}sz line. Furthermore, we give new rigorous bounds  that are asymptotically correct in the limit $h \to 0$ complementing the bounds from  
$\lbrack$J. Ruiz and M. Wouts. On the Kert{\'{e}}sz line: Some rigorous bounds. Journal of Mathematical Physics, 49:053303, May 2008, \cite{Rui}$\rbrack $ , which were asymptotically correct for $h \to \infty$. Finally, using a cluster expansion, we investigate the continuity of the  Kert{\'{e}}sz line phase transition. 
\end{abstract}

\section{Introduction} 
The random-cluster model \cite{Gri06} has been under intense investigation the last 30 years. The model generalises the Fortuin-Kasteleyn graphical representation of the Ising model \cite{FK70} to a graphical representation of all Potts models.

Highlights in the investigations in two dimensions are the calculation of the critical point for all $1 \leq q  < \infty$ in  \cite{BD12}, sharpness of the phase transition \cite{sharpness},  the scaling relations of critical exponents \cite{duminilcopin2020planar} as well as the rigorous determination of the domain of parameters where the phase transition is continuous \cite{DuminilCopin2017ContinuityOT, Diskont2016}. 
In higher dimensions, or in the presence of a magnetic field, results are scarcer with most recent efforts focusing on the near-critical planar regime \cite{CGN14,CJN20, ott, klausen2022mass}.

A magnetic field is implemented in the random-cluster model using Griffith's ghost vertex, which is an additional vertex connected to all other vertices in the graph. The Kert{\'{e}}sz line then separates two regions which are defined according to whether or not there is percolation without using the ghost vertex. The Kert{\'{e}}sz line transition need not correspond to a thermodynamic phase transition (i.e.\ a point where the free energy is not analytic). 
For example, by the Lee-Yang Theorem, the free energy of the Ising model is analytic for all $h \neq 0$ \cite{lee1952statistical, velenik}. Thus, passing through the domain where $h > 0,$ one never encounters a thermodynamic phase transition. Nevertheless, in the random-cluster model, the percolative behaviour may change and thereby signify a geometric phase transition which separates an ordered phase from a disordered phase even at $h>0$. This phase transition defines the  Kert{\'{e}}sz line.  

The Kert{\'{e}}sz line was first studied by  Kert{\'{e}}sz in \cite{Ker} and further discussed in \cite{Rui,Blanchard_2008, CJN18, cusp}. Kert{\'{e}}sz noted that there is no contradiction between the analyticity of the free energy and the geometric phase transition. It was proven in \cite{Blanchard_2008} that for large $q$ and small $h > 0$, where the phase transition is first order, that the Kert{\'{e}}sz line and the thermodynamic phase transition line coincide. 

An interesting feature of the problem is that it involves three variables $(p,q,h)$ and its study has to delve into the trade-off between the decorrelating effect of $h$ on the edges in $\Z^d$, the order-enhancing effect of the inverse temperature governed by the parameter $p$ and the  monotonically decreasing behaviour in $q$.  

\subsection*{Organisation of the paper and main results}
The random-cluster measure on a finite graph $G=(V,E)$ with a distinguished vertex  $\mathfrak{g}$, parameters $p\in (0,1)$ and $q>1$ and external field $h>0$ is the measure on $\{0,1\}^{E}$ given by
$$
\phi_{p,q,h,G}[\{\omega\}]=\frac{1}{Z_{p,q,h,G}}p^{o(\omega_{\text{in}})}(1-p)^{c(\omega_{\text{in}})}p_h^{o(\omega_{\mathfrak{g}})}(1-p_h)^{o(\omega_{\mathfrak{g}})}q^{\kappa(\omega)},
$$
where $\omega_{\mathfrak{g}}$ is the restriction of $\omega$ to the set of edges adjacent to $\mathfrak{g}$, $\omega_{in}$ is the restriction of $\omega$ to the set of edges not adjacent to $\mathfrak{g}$, $o(\cdot)$ denotes the number of open edges and $\kappa(\cdot)$ the number of components. Finally, $p_h=1-\exp(-\frac{q}{q-1}h).$ \\
In our paper, $G$ will be a subgraph of $\mathbb{Z}^d$ with a single external vertex $\mathfrak{g}$ (called the ghost)     added, which is connected to every other vertex. As one fixes two of the three parameters $p,q,h$ and varies the third, the model exhibits a (possibly trivial) percolation phase transition at a point which we denote $p_c(q,h),$ $q_c(p,h)$ and $h_c(p,q)$ respectively. The Kertész line is exactly the set of such points of phase transition. A more precise definition will be given later. We shall often omit one of the variables from notation. In such a case, we consider the omitted variable fixed.

After catching the reader up on some preliminaries, our first order of business in Section 3 is to use the techniques from \cite{grimmett1995comparison} to prove the following:
\begin{theorem} \label{thm:strict_monotonicity}
Let $d \geq 2$. Then, the maps $q \mapsto p_c(q,h),$ $p\mapsto q_c(p,h)$ and $h\mapsto q_c(p,h)$ are strictly increasing and the map $h \mapsto p_c(q,h)$ is strictly decreasing. Furthermore, $q\mapsto h_c(p,q)$ is strictly increasing on $(q_c(p,0),\infty)$ and $p\mapsto h_c(p,q)$ is strictly decreasing on $(p_c(1,0),p_c(q,0))$.
\end{theorem}
Continuity follows from the strict monotonicity and thus, that the Kert{\'{e}}sz line is aptly named - it is, indeed, a curve. In particular, this proves that $h_c(p) >0$ for all $p \in (p_c(1,0), p_c(q,0))$ as was conjectured in \cite[Remark 4]{CJN18}.
\begin{corollary}
The Kert{\'{e}}sz line $ h \mapsto p_c(q,h)$  (and  $p \mapsto h_c(p,q)$) is continuous.
\end{corollary} 
\begin{proof}
Suppose that $h\mapsto p_c(q,h)$ were strictly decreasing but discontinuous at $h_0$. Denote by $p_c^+=\sup_{h>h_0} p_c(q,h)$ and by $p_c^-=\inf_{h<h_0} p_c(q,h)$. Then, we see that for $p\in (p_c^+,p_c^-),$ there is no percolation at $(p,h)$ for $h<h_0,$ but there \textit{is} percolation for $h>h_0$. Accordingly, $p\mapsto h_c(p,q)$ would be constant on $(p_c^-,p_c^+).$ Thus, the corollary follows by contraposition.
\end{proof}

Then, in Section \ref{bounds}, we obtain new upper and lower bounds on the Kert{\'{e}}sz line improving those of \cite{Rui}. In particular, we obtain bounds that asymptotically tend to the correct value in the limit $h \to 0$ complementing the bounds from \cite{Rui}, which asymptotically matched the correct value for $h \to \infty$. 
\begin{theorem} \label{thm:upper_bound} 
There exists an explicit function $\arctanh_q$ such that, for any $q \in \lbrack 1, \infty) $ and any $d\geq 2$, it holds that 
\begin{align*}
h_c(p) \leq  \operatorname{arctanh}_q \left( \sqrt{\frac{1}{q-1} \left(\frac{q}{q_c(p,0)} -1 \right) } \right).  
\end{align*} 
In particular, in the special case of the planar Ising model $q = 2,$ this reduces to 
\begin{align*}
h_c(p) \leq \operatorname{arctanh} \left( \sqrt{\frac{2}{q_c(p,0)} -1} \right)  = \operatorname{arctanh} \left( \sqrt{\frac{2(1-p)^2}{p^2} -1} \right). 
\end{align*}
\end{theorem} 
 Our methods include a generalisation of a recent lemma from \cite{CJN20} concerning the probability of clusters in $\omega_{\text{in}}$ connecting to the ghost which allows us to obtain a stochastic domination result. This allows to bound the model at $(p,q,h)$ from below by the model at $(p,q'(h),0)$ for some explicit value of $q'(h)$. The game is then to make $h$ large enough that $q'(h)<q_c(p),$ so that we achieve percolation. We note that our results here transfer to improve results on the random field Ising model, which was analysed using the  Kert{\'{e}}sz line in \cite{CJN18}.\\ 
In order to get a lower bound for the Kertész line, we employ techniques from \cite{Renormalization} to obtain the following:  Set $\mu:= \frac{(2d+1)^{2d+1}}{(2d)^{2d}}$ and $\Lambda_k=[-k,k]^d\cap \mathbb{Z}^d$.

\begin{theorem} \label{thm:lower_bound}
Suppose that $p < p_c(q,0)$ and that $\delta = \mu^{-4^d}$ and let $k$ be the smallest $k$ such that 
$$
\phi^1_{p,q,0,\Lambda_{3k}}[\Lambda_k\cc \partial \Lambda_{3k}]<\frac{\delta}{2}. 
$$
Then, there is no percolation at $(p,h)$ for
$$
p_h<1-\left(1-\frac{\delta}{2}\right)^{1/|\Lambda_{3k}|}.
$$
\end{theorem} 

In Section \ref{continuity} , we adapt the cluster expansion to this particular setup and use it to prove that the phase transition across the Kert{\'{e}}sz line is continuous when $h$ is sufficiently large. This complements the results of \cite{Blanchard_2008}, where the Pirogov-Sinai theory was used to prove that the phase transition is discontinuous for sufficiently large $q$ and sufficiently small $h >0$. More precisely, let 
$\mathfrak{Z}=\lim_{n\to\infty} \frac{\log(Z_{\Lambda_n})}{|\Lambda_n|}$ denote the pressure. We then prove the following:
\begin{theorem} \label{analytic pressure} There is a function $h_0(q,d)$ such that for $h>h_0(q,d),$
$\mathfrak{Z}$ is an analytic function of $p$. Explicitly,
$$
h_0(q)=\begin{cases}
\left(1+\frac{1}{q-1} \right)^{-1}(\log(2)+\log(q-1)+2d+(2d+1)\log(2d+1)-2d\log(2d)) & q\geq 2 \\
\left(1+\frac{1}{q-1}\right)^{-1}(\log(q)+2d+(2d+1)\log(2d+1)-2d\log(2d)) & q\in (1,2).
\end{cases}
$$
\end{theorem}

We have plotted the function $h_0$ in Figure \ref{continuity_sketch}.

Finally, in Section \ref{outlook}, we provide an outlook on further questions on the Kert{\'{e}}sz line. We discuss continuity and to what extent the Kert{\'{e}}sz line always coinsides with the line of maximal correlation length. Finally, we briefly other models, namely the Loop-$O(1)$ and random current models, for which there is a natural notion of a Kert{\'{e}}sz line, but which lack the sort of monotonicity which is crucial to the study of the random-cluster model. \\

The aim of this paper is two-fold. On the one hand, we aim to answer questions about the Kert{\'{e}}sz line of the Ising model. On the other, we try to provide a unifying  account of the problem, a display of the many different (at times, rather standard) techniques from around the field of percolation that may aid in attacking the problem.

\section{Preliminaries}
We first introduce the Potts model, of which the random-cluster model is a graphical representation. We follow the notation of \cite{DC17} which we also refer to for further information. 

\subsection*{Potts and random-cluster models}
For some finite graph $G = (V,E)$ the Ising model is a probability distributions on the configuration space $ \Omega = \{-1, +1\}^V$. The $q$-state Potts model is a generalisation on the configuration space $(\mathbb{T}^q)^V,$ where $\mathbb{T}^q$ is defined as follows:

\begin{definition}
We define $\mathbb{T}^q$ to be the vertices of a $q$-simplex in $\mathbb{R}^{q-1}$ containing $\mathbf{1}:=(1,0,0,0,...,0)$ as a vertex and such that 
$$
\langle x,y \rangle =\begin{cases} 1 & x=y \\ -\frac{1}{q-1} & else \end{cases}
$$ 
for all vertices $x$ and $y$ of the simplex. 
\end{definition} 
Note that one recovers the spin space $\{-1, +1\}$ of the Ising model for $q=2$. Throughout, for a subgraph $G=(V,E)$ of some graph $\mathbb{G}=(\mathbb{V},\mathbb{E})$, we shall denote by $\partial_e G$ the edge boundary of $\mathbb{G},$ i.e.\ the set of edges in $\mathbb{E}$ with one end-point in $V$ and one end-point outside it. Similarly, the vertex boundary $\partial_v G$ of $G$ is the set of vertices in $V$ incident to edges in $\partial_e G$. 
The Potts Hamiltonian with boundary condition $b$, where $b=0$ corresponds to free boundary conditions and $b=1$ corresponds to boundary condition of all spins pointing in the same direction, is defined as follows.

\begin{definition}
For a finite subgraph $G=(V,E)$ of an infinite graph $\mathbb{G}$, $b\in\{0,1\}$, and $\sigma\in \mathbb{T}_q^{V},$ we define the Hamiltonian
$$
\mathcal{H}^b(\sigma)=- \left(\sum_{e=(i,j)\in E} \langle \sigma_i,\sigma_j\rangle+\id_{\{b=1\}}\sum_{\substack{e=(i,j)\in \partial_e G\\ i\in V}} \langle \sigma_i, \mathbf{1}\rangle\right)
$$  
We define the $q$-state Potts Model partition function with boundary condition $b\in \{0,1\}$, inverse temperature $\beta$, and external field $h$ as 
$$
Z^{b,q}_{\beta,h}(G)=\sum_{\sigma\in \mathbb{T}_q^{V}} e^{-\beta\mathcal{H}^b(\sigma)+h\sum_i \langle \sigma_i, \mathbf{1}\rangle}
$$
The probability of a configuration $ \sigma \in \mathbb{T}_q^V$ is then given by 
$$
\mu^{b,q}_{\beta,h}[\{\sigma\}] =  \frac{1}{Z^{b,q}_{\beta,h}(G)} e^{-\beta\mathcal{H}^b(\sigma)+h\sum_i \langle \sigma_i, \mathbf{1}\rangle}. 
$$
\end{definition}

The random-cluster model is a graphical representation of the Potts model which is of independent interest. In particular, its correlation functions define an interpolation between those of the Potts model for non-integer $q$. 

\begin{definition} For a finite subgraph $G=(V,E)$ of an infinite graph $\mathbb{G}$, $q\geq 1$, vector $\boldsymbol{p}= \{p_e \}_{e \in E}  \in [0,1]^E$ and partition $\xi$ of $\partial_v G$, the random-cluster measure on $G$ with boundary condition $\xi$ is the probability measure $\phi^{\xi}_{\boldsymbol{p},q,G}$ on $\{0,1\}^E$ which, to each $\omega \in \{0,1\}^E$, assigns the probability
$$
\phi^{\xi}_{\boldsymbol{p},q,G}[\{\omega\}]=\frac{1}{Z_{\boldsymbol{p},q,G}^{\xi}}q^{\kappa^{\xi}(\omega)}\prod_{e\in E} p_e^{\omega(e)} (1-p_e)^{1-\omega(e)},
$$
where $\kappa^{\xi}$ is the number of connected components in the graph $$
G^{\xi}_{\omega}=(V, E_{\omega})/\sim_{\xi},
$$
 $E_{\omega}:=\{ e\in E|\;\omega(e)=1\},$ $\sim_{\xi}$ is the equivalence relation with equivalence classes given by $\xi$ and $Z_{\boldsymbol{p},q,G}^{\xi}$ is a normalising constant called the partition function.\\
 $E_{\omega}$ is called the set of \textbf{open edges} and dually, $E\setminus E_{\omega}$ is called the set of \textbf{closed edges}.
\end{definition}
\begin{remark} Two boundary conditions are of special interest. These are the \textit{wired} respectively \textit{free} boundary conditions denoted as $\xi=1$ and $\xi=0$ respectively. $\xi=1$ corresponds to the trivial partition where all boundary vertices belong to the same class and $\xi=0$ corresponds to the trivial partition where all boundary vertices belong to distinct classes.
\end{remark} 
\begin{proposition}[Domain Markov Property]
If $G_1=(V_1,E_1)\subseteq G_2=(V_2,E_2)$ are two finite subgraphs of an infinite graph $\mathbb{G}$, we write $\omega_1:=\omega|_{E_1}$ and $\omega_2:=\omega \mid _{E_2\setminus E_1}$. Then,
$$
\phi_{\boldsymbol{p},q,G_2}^{\xi}[\omega_1\in A|\;\omega_2]=\phi_{\boldsymbol{p},q,G_1}^{\xi_{\omega_2}}[A]
$$
where $v$ and $w$ belong to the same element of $\xi_{\omega_2}$ if and only if they are connected by a path (that might possibly have length $0$) in $((V_2\setminus V_1), E_{\omega_2})/\sim_\xi$.
\end{proposition}
\begin{definition}
For an infinite graph $\mathbb{G}=(\mathbb{V},\mathbb{E})$, we say that a probability measure $\phi_{\boldsymbol{p},q}$ on $\{0,1\}^{\mathbb{E}}$ is an \textbf{infinite-volume random-cluster measure} on $\mathbb{G}$ if, for any finite subgraph $G=(V,E)$ of $\mathbb{G}$, we write $\omega_1=\omega|_E$ and $\omega_2=\omega|_{E^c}$ and have
$$
\phi_{\boldsymbol{p},q}[\omega_1\in A|\; \omega_2]=\phi_{\boldsymbol{p},q,G}^{\xi_{\omega_2}}[A],
$$
where $v$ and $w$ belong to the same element of $\xi_{\omega_2}$ if and only if they are connected by a path in $(\mathbb{V}\setminus V,E_{\omega_2})$.
\end{definition}
\begin{remark} Two natural infinite volume measures occur as monotonic limits of finite volume ones. For an increasing sequence $G_n=(V_n,E_n)$ with $\mathbb{G}=\cup_{n=1}^{\infty} G_n$, we define
\begin{align*}
\phi^{1}_{\boldsymbol{p},q,\mathbb{G}}[A] &=\lim_{n\to\infty} \phi^1_{\boldsymbol{p},q,G_n}[A]\\
\phi^{0}_{\boldsymbol{p},q,\mathbb{G}}[A] &=\lim_{n\to\infty} \phi^0_{\boldsymbol{p},q,G_n}[A]
\end{align*}
for all increasing\footnote{In the interest of the flow of the article, we have postponed the formal definition to Definition \ref{increasing} below.} events $A$ depending only on finitely many edges (so that the probabilities on the right-hand side are well-defined eventually). It is easy to check tha tthe limit does not depend on the choice of sequence $G_n$. Since such events are intersection-stable and generate the product $\sigma$-algebra of $\{0,1\}^{\mathbb{E}}$, this determines the two (possibly equal) infinite volume measures uniquely. That these limits define probability measures is a standard consequence of Banach-Alaoglu, since $\{0,1\}^\mathbb{E}$ is compact. 
\end{remark}
For our purposes, the infinite graph $\mathbb{G}=(\mathbb{V},\mathbb{E})$ will always be an augmented version of $\mathbb{Z}^d$ where we add an extra vertex $\mathfrak{g},$ called the \emph{ghost vertex}, such that the set of vertices for each $d\geq 2$ becomes $\mathbb{V}=\mathbb{Z}^d\cup \{\mathfrak{g}\}$
and the set of edges is
$$
\mathbb{E}=\{(x,y)\in \mathbb{Z}^d| \; \|x-y\|_{\infty}=1\}\cup\{(x,\mathfrak{g})|\; x\in \mathbb{Z}^d\},
$$
with the former set denoting the usual nearest neighbour edge set $\mathbb{E}_d$ of $\mathbb{Z}^d,$ which we shall refer to as the \textit{inner edges}, and the latter denoting the so-called \emph{ghost edges} $\mathbb{E}_{\mathfrak{g}}$ after Griffiths (see, for example, \cite{DC17}).\\
Similarly, we will work with the implicit assumption that a finite subgraph $G=(V,E)$ of $\mathbb{G}$ always has edge sets of the form $$
E=\{(x,y)\in \mathbb{E}|\; x,y\in V\}, 
$$
i.e.\ $G$ is the subgraph induced by $V$. 

This also gives a natural partition of $E=E_{\win}\cup E_{\mathfrak{g}}$ similar to that of $\mathbb{E}$. \\
One class of finite subgraphs of special interest is the class of boxes 
$$
\Lambda_k(v)=\{w\in \mathbb{Z}^d|\; \|v-w\|_{\infty}\leq k\} \cup \{ \mathfrak{g} \},
$$
for a fixed vertex $v\in \mathbb{V}$.\\
 The connection of the random-cluster model to the Potts Model goes through the Edwards-Sokal coupling (see, for instance, \cite{DC17}). For any finite subgraph $G$ of $\mathbb{G}$, we have
 $$
 Z_{\boldsymbol{p},q,G}^{b}=e^{-\beta |E|-h|V|}Z_{\beta,h, G}^{b,q}
 $$
 for the specific choice of edge parameters
\begin{align} \label{edge weights}
p_e=\begin{cases} 
p:=1-\exp(-\frac{q}{q-1}\beta) & e\in \mathbb{E}_d \\ p_h:=1-\exp(-\frac{q}{q-1}h) & e\in \mathbb{E}_{\mathfrak{g}}. 
\end{cases}
\end{align}
In keeping with this, we will simply write measure
$$\phi^{\xi}_{p,q,h,G}[\{\omega\}]= \frac{1}{Z_{p,q,h,G}^{\xi}} p^{o(\omega_{in})}(1-p)^{c(\omega_{in})}p_h^{o(\omega_{\mathfrak{g}})}(1-p_h)^{c(\omega_{\mathfrak{g}})},
$$ with $o(\omega_{in})$, respectively $c(\omega_{in})$, denoting the number of open, respectively closed edges, in $E_{\win}$ - and similarly for the ghost edges.
\begin{remark} The case $q=1$, henceforth called Bernoulli percolation, is somewhat particular. Here, the state of the edges becomes a product measure and \eqref{edge weights} no longer gives a translation between an external field strength $h$ and an edge parameter $p_h$. As such, for Bernoulli percolation, we shall instead directly write 
$$
\mathbb{P}_{p,p_h,G}:=\phi_{\boldsymbol{p},1,G},
$$
where $p_e=p$ for $e\in E_{\win}$ and $p_e=p_h$ for $e\in E_{\mathfrak{g}}$. \\
Note that, by independence, $\mathbb{P}_{p,p_h,G}$ is the same for any boundary condition and thus, we drop it from the notation.
\end{remark}
Throughout, we shall think of $h$ as an enhancing parameter boosting the number of interior edges rather than being an edge parameter. This is because the percolation properties of the ghost vertex $\mathfrak{g}$ itself are trivial. In other words, when studying the random-cluster model, if $\omega\sim \phi_{p,q,h,\mathbb{G}}$, we are interested in the percolation phase transition of the marginal $\omega_{in}:=\omega|_{\mathbb{E}_d}$, the distribution of which we will denote simply by $\phi_{p,q,h,\mathbb{Z}^d}$. More precisely, we define the critical parameter $p_c=p_c(q,h,d)$ as
$$
p_c:=\inf\{p |\; \theta(p,q,h)>0\},
$$
where $\theta(p,q,h)=\phi^{1}_{p,q,h,\mathbb{Z}^d}[0 \leftrightarrow \infty]$ 
is the probability that $0$ is part of an infinite connected component of inner edges (the cluster of the ghost vertex $\mathfrak{g}$ is trivially infinite almost surely for $h > 0$). When $\theta(p,q,h)>0$ (and correspondingly, an infinite cluster exists almost surely), we say that the model \emph{percolates}.
 
 The random-cluster model has many nice properties. First of all, it is a graphical representation of the Potts model in the sense that for all $x,y \in V$ of some subgraph $G$ of $\mathbb{G}$, whether finite or infinite, (see \cite[(1.5)]{DC17}),
 \begin{align*}
  \phi_{p,q,h, G}^b[x \cc y] = \mu^{b,q}_{\beta,h,G}[\sigma_x\cdot \sigma_y], 
 \end{align*} 
where $x\cc y$ denotes the event that there is an open path connecting $x$ and $y$ in $(V,E_{\omega})$ and $b\in \{0,1\}$.
For any graph $G=(V,E)$ (finite or infinite),  there is a natural partial order $ \preceq $ on the space of percolation configurations $\{0,1 \}^{E}$. We say that $ \omega \preceq  \omega' $ if $\omega(e) \leq \omega'(e)$ for all $e \in  E \cup E_\mathrm{g}$. This furthermore gives us a notion of events which respect the partial order.
\begin{definition} \label{increasing} We call an event $A$ \emph{increasing} if, for any $\omega \in A,$ it holds that $\omega \preceq \omega'$ implies $\omega' \in A$.\\
 For two percolation measures $\phi_1 ,\phi_2$ we say that $\phi_1$ is \emph{stochastically dominated} by $\phi_2$ if  $\phi_1[A] \leq  \phi_2[A]$ for all increasing events $A$. We will also denote this order relation as $\phi_1  \preceq  \phi_2$. 
 \end{definition}
 The study of increasing events turns out to be natural due to the fact that they are all positively correlated (see \cite{DC17}). 
 \begin{proposition}[FKG] \label{FKG}
 For any two increasing events $A, B,$ any $(p,q,h),$ any boundary condition $\xi$ and graph $G$, we have
 $$
 \phi_{p,q,h,G}^{\xi}[A]\phi_{p,q,h,G}^{\xi}[B] \leq  \phi_{p,q,h,G}^{\xi}[A\cap B].
 $$
 In particular, $\phi_{p,q,h,G}^{\xi} \preceq \phi_{p,q,h,G}^{\xi}(\cdot |A)$.
 \end{proposition}
Furthermore, the random-cluster model carries many other natural monotonicity properties in the sense of stochastic domination (see \cite{DC17}), making it a natural dependent percolation processes to study.  

\begin{theorem} \label{comparison theorem}
For the random-cluster model on some subgraph $G$ of $\mathbb{G}$ (finite or infinite), the following relations hold:
\begin{enumerate}
\item[$i)$] $\phi^{\xi}_{p_0,q_0,h_0,G}$ is monotonic in $\xi$ in the sense that if $\xi'$ is a finer partition than $\xi$, we have
$$
\phi_{p,q,h, G}^{\xi'}  \preceq   \phi_{p,q,h,G}^{\xi} 
$$
for any parameters $(p,q,h)$.
\item[$ii)$] $\phi$ is increasing in $p$, $h$ and decreasing in $q$, i.e.\ if $p'\geq p$ and $h'\geq h$ and $q'\leq q$, we have
$$
\phi^{\xi}_{p,q,h,G}\preceq \phi^{\xi}_{p',q',h',G}
$$
for any boundary condition $\xi$.
\item[$iii)$] The random-cluster model is comparable to Bernoulli percolation in the following sense: for $p_e$ still given by \eqref{edge weights}, we have
$$
\mathbb{P}_{\tilde{p},\tilde{p}_h,G}\preceq \phi^{\xi}_{p,q,h,G} \preceq \mathbb{P}_{p,p_h,G} 
$$
for any boundary condition $\xi$, where $\tilde{p}=\frac{p}{p+q(1-p)}$ and $\tilde{p}_h=\frac{p_h}{p_h+q(1-p_h)}$.
\end{enumerate}
\end{theorem} 

It follows from $iii)$ that for $d \geq 2, q \geq 1, h\geq 0,$ the random-cluster model always has a non-trivial phase transition,  meaning that $p_c\in (0,1)$. \\ 
Furthermore, any of these stochastic domination relations can be realised as a so-called \textit{increasing coupling}. That is, for any two of the above above measures $\mu$ and $\nu$, if $\mu\preceq \nu$, there exists a measure $\mathbb{P}$ on $\{(\omega_1,\omega_2)|\; \omega_1,\omega_2\in \{0,1\}^E\}$ with the property that $\mathbb{P}[\omega_1\preceq \omega_2]=1$ and for any event $A$,
\begin{align*}
\mathbb{P}[\omega_1\in A]&=\mu[A] \\
\mathbb{P}[\omega_2\in A]&=\nu[A]
\end{align*}
For an explicit construction, see \cite[Lemma 1.5]{DC17}.
\subsection*{Definition of the Kert{\'{e}}sz line}  

In the following we introduce the Kert{\'{e}}sz line  following \cite{CJN18}, see also \cite{Rui,Ker}. 
\begin{definition}
 Suppose that $q$ and $d$ are fixed. Then the Kert{\'{e}}sz line is defined by 
 \begin{align*}
h_c(p) = h_c(p,q,d)= \sup\{ h \geq 0 \mid \theta(p,q,h) = 0 \}, 
\end{align*}
i.e.\ $h_c(p)$ is the largest $h$ such that $p\leq p_c(q,h,d)$. 
\end{definition}

The facts that $\{ 0 \cc \infty \}$ is an increasing event and $\phi_{p,q,h,\mathbb{Z}^d}^1$ is monotone in $p$ imply that the Kert{\'{e}}sz line is monotonically decreasing in $p$. 
Furthermore, translation invariance as well as monotonicity of  $\phi_{p,q,h,\mathbb{Z}^d}^1$ in $h$ implies that the Kert{\'{e}}sz line separates $\{(p,h) \mid p \geq 0, h \geq 0 \}$ into two regions with and without an infinite cluster of inner edges. 
If $p > p_c:=p_c(q,0,d),$ then monotonicity in $h$ implies that $h_c(p)=0$. Stochastic domination of the random-cluster model by Bernoulli bond percolation (see Theorem \ref{comparison theorem}) implies that if $p <p_B:= p_c(1,0,d),$  then $h_c(p) = \infty$  (see \cite{CJN18}, Section 3). 

In \cite[Theorem 7-8]{CJN18}, it is proven that if  $p \in (p_B, p_c),$ then $ 0 < h_c(p) < \infty$ except for the case where $p$ is close to $p_c$ where strict positivity of the Kert{\'{e}}sz line is not proven. This will be proven in the next section, thereby settling the question of whether the Kert{\'{e}}sz line for the random-cluster model on $\Z^d$ is always non-trivial. 

\subsection*{Probability of connecting to the ghost given configuration of inner edges} 
In the following, we generalise \cite[Lemma 2.4]{CJN20}, which is stated only in the case of FK-Ising (i.e.\ $q=2$), to the setting of the  random-cluster model for $q \in \lbrack 1, \infty)$. This generalisation, which is straight forward, was first noted in the appendix of \cite{camia2020fk}.
The lemma and some of the bounds that follow from it are more easily stated if we define the function $\tanh_q : \lbrack 0, \infty)  \to \lbrack 0, 1 )$ by  
\begin{align*}
\tanh_q(x) = \frac{1-e^{-2x}}{(q-1)e^{-2x} +1}.
 \end{align*} 
 Notice that $\tanh_2 = \tanh $, that $\tanh_q$ is strictly increasing, satisfies $\tanh_q( \lbrack 0, \infty) ) = \lbrack 0,1)$ and has an inverse
 which we call $\text{arctanh}_q: \lbrack 0,1) \to \lbrack 0, \infty)$ given by 
 \begin{align*}
\arctanh_q(x) = \frac{1}{2}  \log \left( \frac{(q-1)x +1}{1-x} \right). 
 \end{align*}
Then, we can state the following lemma.
\begin{lemma} \label{ghosted}
	Suppose that $G$ is a finite graph and that a configuration of inner edges $\omega_{in}$ has clusters $C_1 , \dots, C_n$. Then, for each $1\leq i \leq n,$  
	\begin{align*}
	\phi^0_{p,q,h,G}[ C_i \leftrightarrow \mathfrak{g} \; \vert \;\omega_{\win} ] = \tanh_q(h \abs{C_i})
	\end{align*}
	and the events $\{ C_i \leftrightarrow \mathfrak{g} \}$ are mutually independent given $\omega_{\win}$ . 
\end{lemma}
\begin{proof}
	First, we look at inner edges and compute
	\begin{align*}
	Z^0_{p,q,h,G}\phi^0_{p,q,h,G}[\{ \omega_{\win} \}]  = p^{o(\omega_{\win} )} (1-p)^{c(\omega_{\win})}q \prod_{j=1}^n  \left( \sum_{f \in \{0,1\}^{C_j}} p_h^{o(f)} (1-p_h)^{c(f)} (q \id_{ \{ C_j \not \cc \mathfrak{g} \}} +  \id_{ \{ C_j \cc \mathfrak{g} \}})  \right) 
	\end{align*}
	with $o(f)$ respectively $c(f)$ denoting the number of open respectively closed ghost edges in $C_i$.  From this decomposition, conditional independence follows. Now, to obtain the desired form, note that we get a factor of $q$ from the cluster $C_i$ if and only if all the edges are closed and thus, we can write the sum as 
	\begin{align*}
	 & \sum_{f \in \{0,1\}^{C_i}} p_h^{o(C_i(f))} (1-p_h)^{c(C_i(f))} (q \id_{ \{ C_i \not \cc \mathfrak{g} \}}) +  \id_{ \{ C_i  \cc \mathfrak{g} \}}) \\
	& = q (1-p_h)^{\abs{C_i}} + \sum_{f \in \{0,1\}^{C_i}} p_h^{o(C_i(f))} (1-p_h)^{c(C_i(f))} - (1-p_h)^{\abs{C_i}} \\
	& =  (q-1)(1-p_h)^{\abs{C_i}}  + 1  =  (q-1)e^{ -2 h  \abs{C_i} } + 1. 
	\end{align*} 
	This then means that 
	\begin{align*}
	Z^0_{p,q,h,G}\phi^0_{p,q,h,G}[\{ \omega_{\win} \}] =  p^{o(\omega_{in})} (1-p)^{c(\omega_{\win})} q\prod_{j=1}^n \left((q-1) e^{ -2 h \abs{C_j} } + 1 \right).
	\end{align*}
	Then, we are ready to compute 
	\begin{align*}
	\phi^0_{p,q,h,G}[C_i \leftrightarrow \mathfrak{g} \; \vert\;  \omega_{\win} ] = \frac{   \phi^0_{p,q,h,G }[ \{ C_i \leftrightarrow \mathfrak{g} \}  \cap \{ \omega_{\win} \} ] }{ \phi^0_{p,q,h,G}[\{ \omega_{\win} \}]} 
	= \frac{1 -  e^{ -2 h \abs{C_i}} }{ (q-1) e^{-2 h \abs{C_i}}  + 1 } = \tanh_q( h \abs{C_i}),
	\end{align*}
	which proves the formula. 
\end{proof}
Since the lemma is true uniformly in finite volume it transfers to the infinite volume limit.

\section{Strict monotonicity} \label{strict_monotonicity} 
\newtheorem{prop}{Proposition}

We prove that the Kert{\'{e}}sz line is strictly monotone in all of its parameters. All the following results are uniform in the boundary conditions, so we shall suppress them for ease of notation. The proofs all follow the strategy from \cite{grimmett1995comparison} rather closely. 

In doing so, we will need the following elementary lemma:
\begin{lemma} \label{Sampling algo}
Let $G=(V,E)$ be a finite graph, $F$ some finite, totally ordered set, $\nu$ a probability measure on $\{\eta:E\to F\}$ and $(U_j)_{1\leq j\leq |E|}$ a family of i.i.d. uniform random variables. For any enumeration $(e_j)_{1\leq j \leq |E|}$ of $E$, if we recursively define
\begin{align*}
X(e_1)&=\min\left\{ f\in F\Big| \; U_1\leq \sum_{g\leq f} \nu[\eta(e_1)=g]\right\}\\
X(e_{j+1})&=\min\left\{ f\in F\Big| \; U_{j+1}\leq \sum_{g\leq f} \nu[\eta(e_{j+1})=g|\; \eta(e_i)=X(e_i)\; \forall i\leq j]\right\}, 
\end{align*}
then $X\sim \nu$.
\end{lemma}
\begin{proof}
This follows from the Law of Total Probability as soon as we establish that $X(e)$ has the correct marginal for every $e$. For $e=e_1$, we simply see that
$$
\mathbb{P}[X(e_1)=f]=\mathbb{P}\left[\sum_{g<f}\nu[\eta(e_1)=g]<U_1\leq \sum_{g\leq f} \nu[\eta(e_1)=g]\right]=\nu[\eta(e_1)=f].
$$
The same calculation shows that $X(e_{j+1})$ has the correct conditional law given $(X(e_i))_{1\leq i\leq j}$ and thus, the proposition follows by finite induction.
\end{proof}

Next, in order to prove Theorem \ref{thm:strict_monotonicity}, we prove the following technical proposition, which is in the spirit of \cite[Theorem 2.3]{grimmett1995comparison} (where a similar comparison result between $p-$ and $q-$derivatives at $h=0$ is proved). 
\begin{proposition} \label{comparison ineq}
There exist strictly positive smooth functions $\alpha,\gamma: (0,1)\times (0,\infty)\times (1,\infty)\to \mathbb{R}$ such that, for any finite subgraph 
$G$ of $\mathbb{G}$ containing $\mathfrak{g}$ and any increasing event $A$ depending only on the edges of $G$, we have
$$
\alpha(p,q,h)\frac{\partial}{\partial h} \phi_{p,q,h,G}[A]\leq \frac{\partial}{\partial p}\phi_{p,q,h,G}[A]\leq \gamma(p,q,h)\frac{\partial}{\partial h}\phi_{p,q,h,G}[A].
$$
\end{proposition}
\begin{proof}
We are going to give the explicit construction of $\gamma$. The construction of $\alpha$ is analogous.
A direct calculation shows that
\begin{align*}
\frac{\partial}{\partial p} \phi_{p,q,h,G}[A] &=\frac{1}{p(1-p)} \operatorname{Cov}[o(\omega_{\win}), \id_A]\\
\frac{\partial}{\partial h} \phi_{p,q,h,G}[A] &= \frac{1}{p_h(1-p_h)}\operatorname{Cov}[o(\omega_{\mathfrak{g}}) \id_A].
\end{align*}
If $(\omega_1,\omega_2)$ denotes an increasing coupling with marginals $\omega_1\sim \phi_{p,q,h,G}$ and $\omega_2\sim \phi_{p,q,h,G}(\cdot|A)$, we get that
\begin{align*}
\operatorname{Cov}[o(\omega_{\win}), \id_A] &=\phi_{p,q,h,G}[A]\mathbb{E}[o(\omega_{2,\win})-o(\omega_{1,\win})]\\
\operatorname{Cov}[o(\omega_{\mathfrak{g}}), \id_A]&=\phi_{p,q,h,G}[A]\mathbb{E}[o(\omega_{2,\mathfrak{g}})-o(\omega_{1,\mathfrak{g}})].
\end{align*}
Thus, we are finished if we can establish that $\mathbb{E}[o(\omega_{2,\win})-o(\omega_{1,\win})]$ and $\mathbb{E}[o(\omega_{2,\mathfrak{g}})-(\omega_{1,\mathfrak{g}})]$ are comparable.

This, however, turns out to be easy, because
\begin{align} \label{sum_it_up}
\mathbb{E}[o(\omega_{2,\win})-o(\omega_{1,\win})] &=\sum_{e\in E_{\win}} \mathbb{P}[\omega_2(e)=1,\omega_1(e)=0] \\
\mathbb{E}[o(\omega_{2,\mathfrak{g}})-o(\omega_{1,\mathfrak{g}})] &=\sum_{e\in E_{\mathfrak{g}}} \mathbb{P}[\omega_2(e)=1,\omega_1(e)=0].
\end{align}
Throughout, for $e\in E$, we let $B_e$ denote the event $[\omega_2(e)=1,\omega_1(e)=0]$. For $v\in V\setminus \{\mathfrak{g}\},$ let $B_v$ denote the event that all neighbouring edges of $v$ in $\mathbb{Z}^d$ are closed in $\omega_1$ and $\mathfrak{G}_v$ denote the event that the ghost edge $(v,\mathfrak{g})$ is open in $\omega_1$.

If $v$ and $w$ are the two end-points of $e$, we claim that
$$
\mathbb{P}[B_v\cap \mathfrak{G}_{w}|B_e]\geq (1-p)^{2d-1}\frac{p_h}{p_h+q(1-p_h)}.
$$
To see this, we apply Lemma \ref{Sampling algo} with $F=\{(0,0),(1,0),(1,1)\}$ (with the obvious ordering) and any enumeration such that $e_1=e$, $e_2=(w,\mathfrak{g})$ and $e_3,...,e_{2d+1}$ are some enumeration of the other neighbours of $v$. Allowing ourselves the slight abuse of notation of keeping our letters, we get that
$$
\mathbb{P}[B_v\cap \mathfrak{G}_{w}|B_e]\geq \mathbb{P}\left[\left\{U_2\geq \frac{p_h}{p_h+q(1-p_h)}\right\}\cap \bigcap_{j=3}^{2d+1}\{U_j\leq (1-p)\}\Bigg|\; B_e\right],
$$
where we have used $iii)$ in Theorem \ref{comparison theorem} to bound the conditional distribution of $\omega_1$ from above and from below by Bernoulli percolation. To establish the above claim, we simply note that $B_e$ is measurable with respect to the $\sigma$-algebra generated by $U_1$.

Now, on the event $B_v \cap \mathfrak{G}_{w}\cap B_e$, the end-points of $(v,\mathfrak{g})$ are disconnected in  $\omega_1$ and so, we get 
$$
\mathbb{P}[\omega_1((v,\mathfrak{g}))=1|B_e\cap B_v\cap \mathfrak{G}_{w}]=\phi^0_{p,q,h,\{v,\mathfrak{g}\}}[(v,\mathfrak{g})]=\frac{p_h}{p_h+q(1-p_h)},
$$
by applying Lemma \ref{Sampling algo} in a similar fashion.

 Meanwhile, the end-points are connected in $\omega_2$, so that 
 $$\mathbb{P}[\omega_2((v,\mathfrak{g})=1)|B_e\cap B_v \cap \mathfrak{G}_{w}]\geq \phi^1_{p,q,h,\{v,\mathfrak{g}\}}[(v,\mathfrak{g})]=p_h,
 $$
where we have used the FKG-inequality (Proposition \ref{FKG}). All in all, we get that
 $$
 \mathbb{P}[B_{(v,\mathfrak{g})}|B_e]\geq \left(p_h-\frac{p_h}{p_h+q(1-p_h)}\right)\frac{p_h}{p_h+q(1-p_h)}(1-p)^{2d-1}:=\gamma'
 $$
 Thus, for every inner edge $e$, pick an end-point $v(e)$.
Since any given vertex $v$ belongs to at most $2d$ different inner edges, we get that
 \begin{align*}
 \gamma' \mathbb{E}[o(\omega_{2,\win})-o(\omega_{1,\win})] &\leq \sum_{e\in E_{\win}} \mathbb{P}[B_{(v(e),\mathfrak{g})}|B_e]\mathbb{P}[B_e]
 \leq 2d\sum_{v\in V} \mathbb{P}[B_{(v,\mathfrak{g})}]
 \leq 2d \sum_{e\in E_{\mathfrak{g}}} \mathbb{P}[B_e]. 
 \end{align*}
Defining $\gamma:=2d\frac{p_h(1-p_h)}{\gamma' p(1-p)}$ completes the construction.
\end{proof}

 \begin{figure}
\captionsetup{width=.6\linewidth}
{  \caption{Sketch of the event $B_v \cap \mathfrak{G}_{w}\cap B_e$ from the proof of Proposition \ref{comparison ineq}. With dotted lines we denote closed edges and with solid lines open edges. The edge in question $(v,\mathfrak{g})$ is not depicted. The probability that it is open differs in the two cases $\omega_1$ and $\omega_2$.    \label{monotonicity_sketch}}}{
     \includegraphics[scale =0.2]{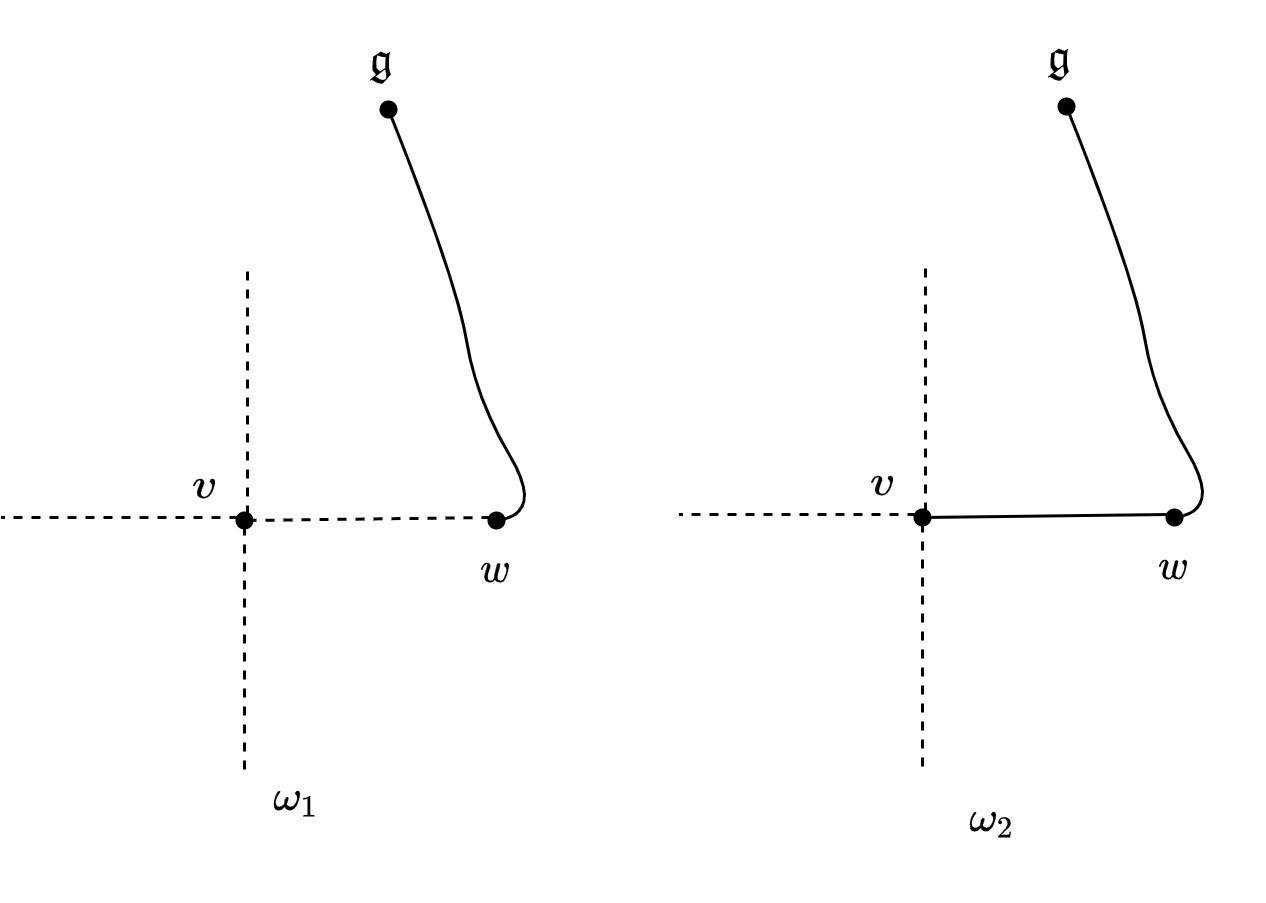}}
\end{figure}

Meanwhile, the following is a direct analogue of \cite[Theorem 2.3]{grimmett1995comparison} for the case with $h>0$.
\begin{proposition}
There exist strictly positive smooth functions $\alpha,\gamma: (0,1)\times (0,\infty)\times (1,\infty)\to \mathbb{R}$ such that, for any finite subgraph
$G$ of $\mathbb{G}$ containing $\mathfrak{g}$ and any increasing event $A$ depending only on the edges of $G$, we have
$$
-\alpha(p,q,h) \frac{\partial}{\partial q} \phi_{p,q,h,G} 
 [A]\leq \frac{\partial}{\partial p} \phi_{p,q,h,G}[A]+\frac{\partial}{\partial h} \phi_{p,q,h,G}[A]\leq -\gamma(p,q,h) \frac{\partial}{\partial q} \phi_{p,q,h,G}[A]. 
$$
\end{proposition}
\begin{proof}
Similarly to the previous result, one obtains via a direct calculation that
$$
\frac{\partial}{\partial q} \phi_{p,q,h,G}[A]=\frac{1}{q}\operatorname{Cov}[\id_A,\kappa],
$$
where $\kappa$ denotes the number of components of a configuration (counted according to the appropriate boundary conditions). Once again, letting $(\omega_1,\omega_2)$ denote some increasing coupling of $\omega_1\sim \phi_{p,q,h,G}$ and $\omega_2\sim \phi_{p,q,h,G}[\cdot|A]$, we get
$$
-\operatorname{Cov}[\id_A,\kappa]=\phi_{p,q,h,G}[A]\mathbb{E}[\kappa(\omega_1)-\kappa(\omega_2)].
$$
It is immediate that  
$$\kappa(\omega_1)-\kappa(\omega_2)\leq o(\omega_{2,\win})-o(\omega_{1,\win})+o(\omega_{2,\mathfrak{g}})-o(\omega_{1,\mathfrak{g}}),$$
 which allows one to construct $\alpha$ using (\ref{sum_it_up}) from the proof of the previous proposition.\\
For the other inequality, let $N_I(\omega_1,\omega_2)$ denote the number of vertices which are isolated in $\omega_1$ but not in $\omega_2$. Then, again, since $\omega_1 \preceq \omega_2,$ it holds that
$$\kappa(\omega_1)-\kappa(\omega_2)\geq \frac{1}{2}N_I(\omega_1,\omega_2)=\frac{1}{2}\sum_{v\in V} \id_{I_v},
$$
where $I_v$ is the event that $v$ is isolated in $\omega_1$ but not in $\omega_2$. Just like before, for every edge $e$ incident to $v$, one can argue the existence of some smooth function $\gamma'$ such that
$$
\mathbb{P}[I_v|B_e]\geq \gamma'.
$$
Summing over them all, we get that 
$$
\mathbb{P}[I_v]\geq \frac{\gamma'}{2d}\sum_{e\sim v} \mathbb{P}[B_e], 
$$
Using that any edge is incident to exactly two vertices, we have
$$
\mathbb{E}[\kappa(\omega_1)-\kappa(\omega_2)]\geq \frac{2\gamma'}{4d}\sum_{e\in E}\mathbb{P}[B_e]\geq \frac{\gamma'}{2d}\left(\mathbb{E}[o(\omega_{2,\win})-o(\omega_{1,\win})]+\mathbb{E}[o(\omega_{2,\mathfrak{g}})-o(\omega_{1,\mathfrak{g}})]\right),
$$
from which $\gamma$ may be constructed easily.
\end{proof}

Finally, we are in position to prove the main result. 
\\
\\
\textit{Proof of Theorem \ref{thm:strict_monotonicity} :}We are going to prove that $h\mapsto p_c(q,h)$ is strictly decreasing. The other statements are proven completely analogously.
Consider $q$ fixed and let $\alpha(p,h)$ be the function from Proposition \ref{comparison ineq}. For $(p,h),$ let $\ell_{p,h}=(\ell_{p,h}^1,\ell_{p,h}^2):(T^{-}_{p,h},T^{+}_{p,h})\to (0,1)\times (0,\infty)$ be the integral curve of the vector field $(-1,\alpha)$ started at $(p,h),$ i.e.\ a maximal solution to
\begin{align*}
\frac{d}{dt}\ell_{p,h}(t)&=(-1,\alpha(\ell_{p,h}(t))) \\
\ell_{p,h}(0) &=(p,h).
\end{align*}
The existence of such a solution is completely standard ODE fare. For instance, one may appeal to the Picard-Lindelöf Theorem.

Now, for any finite subgraph $G$ of $\mathbb{G}$ containing $\mathfrak{g}$ and any increasing event $A$, we find that
$$
\frac{d}{dt} \phi_{\ell_{p,h}^1(t),q,\ell_{p,h}^2(t),G}[A]=\frac{d}{dt}\ell_{p,h}(t)\cdot \nabla\phi_{\ell_{p,h}^1(t),q,\ell_{p,h}^2(t),G}[A]\leq 0,
$$
by construction of $\alpha$. Since $A$ was arbitrary, we conclude that $\phi_{\ell_{p,h}^1(t_1),q,\ell_{p,h}^2(t_1),G}\preceq \phi_{\ell_{p,h}^1(t_2),q,\ell_{p,h}^2(t_2),G}$ for $t_1<t_2$. Furthermore, since $G$ was arbitrary, this extends to any infinite volume limits.

Let $h<h'$ be such that $\ell_{p_c(h),h}(t)$ intersects $\mathbb{R}\times\{h'\}$ and denote by $(\tilde{p}, h')$ the point of intersection. Since $\alpha>0,$ the second coordinate of $\ell_{p,h}(t)$ is strictly increasing in $t$ and the first coordinate strictly decreasing. Thus, $\tilde{p}<p_c(h)$.

For any $p>\tilde{p},$ let $(\hat{p},h)$ denote the intersection of $\ell_{p,h'}(t)$ with $\mathbb{R}\times \{h\}$. Then, since the paths of the different integral curves are either identical or non-intersecting, we must have that $\hat{p}>p_c(h)$. Thus, we have that $\phi_{\hat{p},q,h}$ almost surely percolates.

However, $\phi_{\hat{p},q,h}\preceq \phi_{p,q,h'},$ so we conclude that $\phi_{p,q,h'}$ almost surely percolates. We conclude that
$$
p_c(h')\leq \tilde{p}<p_c(h),
$$
which is what we wanted.
\qed

 \begin{figure}
\captionsetup{width=.6\linewidth}
{  \caption{The points $(p_c(h), h), (\tilde p, h'), (p,h'), (\hat p, h)$ as well as the integral curves from the proof of  Theorem \ref{thm:strict_monotonicity}.} \label{integral_curve}}{\includegraphics[scale =0.25]{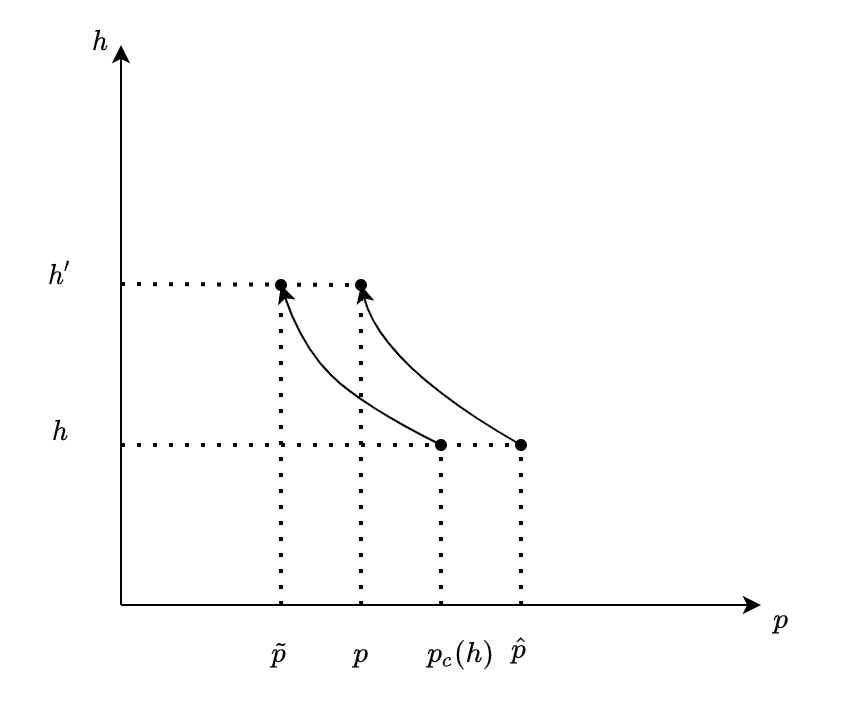}}
\end{figure}


\section{Stochastic Domination and bounds} \label{bounds}
In this section, we investigate conditions for stochastic domination. Since the event $\{ 0 \cc \infty \} $ is increasing, stochastic domination results directly transfer to bounds on the  Kert{\'{e}}sz line.
Here and in the following, let $r_i = \frac{p_i}{1-p_i}$ and $H_i = \frac{{p_h}_i}{1-{p_h}_i} $ for $i \in \{1,2\}$. 
\begin{theorem}\label{eksplicitdom}
Suppose that $0 \leq p_2 \leq p_1 \leq 1, q_i \in \lbrack 1, \infty),h_i \geq 0$ for $i \in \{1,2\}$ and that for all positive integers $n,m$ it holds that
\begin{align*}
\frac{r_2}{q_2} \left( \left(q_2 -1 \right)\tanh_{q_2}(n h_2)\tanh_{q_2}(m h_2 ) + 1  \right) \leq  \frac{r_1}{q_1} \left( \left(q_1 -1 \right)\tanh_{q_1}(n h_1)\tanh_{q_1}(m h_1 ) + 1  \right). 
\end{align*}
Then, for any increasing event $A$ that only depends on the inner edges,  we have
	\begin{align*}
	\phi_{p_2, q_2, h_2}[A] \leq \phi_{p_1, q_1, h_1}[A].
	\end{align*}	
\end{theorem}
Letting  $p_1 = p_2$ and $h_2 =0$ and using that $\tanh_{q_1}$ is increasing, we obtain the following results from which our bounds will follow. 	

\begin{corollary}	
Suppose that $q_1,q_2 \in \lbrack 1, \infty)$ and $h_1 \geq 0$ are such that  
\begin{align*}
    \frac{1}{q_2} \leq  \frac{1}{q_1} \left( \left(q_1 -1 \right)\tanh_{q_1}(h_1)^2 + 1  \right). 
\end{align*}
Then, for any increasing event $A$ that only depends on the inner edges, 
	\begin{align*}
	\phi_{p, q_2,0}[A] \leq \phi_{p, q_1, h_1}[A].
	\end{align*}	    
	
\end{corollary}

In order to obtain the result, we modify an argument from  \cite[Theorem 3.21]{Gri06}.  

\begin{proof}[Proof of Theorem \ref{eksplicitdom}] 
Again, we prove the result uniformly over all finite graphs $G$. This then transfers to any infinite-volume limits.\\
Let $X$ be an increasing random variable which only depends on inner edges. 
Then,
$$
\phi_{p_j, q_j, h_j,G}[X] = \frac{1}{Z_j} \sum_{\omega_{\win}\in \{0,1\}^{E_{\win}}}  r_j^{o(\omega_{\win})} X(\omega_{\win}) \sum_{\omega_{\mathfrak{g}}\in \{0,1\}^{E_{\mathfrak{g}}}} H_j^{o(\omega_{\mathfrak{g}})} q_j^{\kappa(\omega)},
$$
where we have abbreviated $Z_j:=(1-p_j)^{|E|}(1-p_{h_j})^{|V|}Z^{\xi}_{p_j,q_j,h_j,G}$.\\
Now, define the random variable $Y$ depending on inner edges $\omega_{\win}$ by
\begin{align*}
Y(\omega_{\win}) = \frac{r_1^{o(\omega_{\win})}\sum_{\omega_{\mathfrak{g}}} H_1^{o(\omega_{\mathfrak{g}})} q_1^{\kappa(\omega)}}{r_2^{o(\omega_{\win})}\sum_{\omega_{\mathfrak{g}}} H_2^{o(\omega_{\mathfrak{g}})} q_2^{\kappa(\omega)}}. 
\end{align*}

This lets us rewrite
\begin{align*}
\phi_{p_1, q_1, h_1,G}[X] = \frac{Z_2}{Z_1} \frac{1}{Z_2} \sum_{\omega_{\win}} X[\omega_{\win}] Y(\omega_{\win})r_2^{o(\omega_{\win})}\sum_{\omega_{\mathfrak{g}}} H_2^{o(\omega_{\mathfrak{g}})} q_2^{\kappa(\omega)} = \frac{Z_2}{Z_1}  \phi_{p_2, q_2, h_2,G}[XY]. 
\end{align*}
Letting $X =1,$ one gets 
\begin{align*}
1 = \phi_{p_1, q_1, h_1,G}[1] = \frac{Z_2}{Z_1} \sum_{\omega_{\win}}  Y(\omega_{\win})r_2^{o(\omega_{\win})}\sum_{\omega_{\mathfrak{g}}}H_2^{o(\omega_{\mathfrak{g}})}q_w^{\kappa(\omega)},
\end{align*}
from which we can conclude 
\begin{align*}
 \phi_{p_1, q_1, h_1,G}[X] =   \frac{\phi_{p_2, q_2, h_2,G}[XY] }{ \phi_{p_2, q_2, h_2,G}(Y) }. 
\end{align*}
Now, if we can prove that $Y$ is increasing, then by FKG, we get that
$
\phi_{p_2, q_2, h_2,G}[X]   \leq  \phi_{p_1, q_1, h_1,G}[X],
$
since $X$ is increasing. 
Since $X$ was arbitrary, this would prove the stochastic domination between the marginals on $E_{\win}$. \\
Therefore, in the following, we focus on $Y$ and whether it is increasing. Let an edge $e = (x,y)$ be given. Then, the condition for $Y$ to be increasing is
$
Y(\omega_{\win}) \leq Y(\omega_{\win}^e) 
$
where $\omega^e$ denotes the configuration $\omega$ where we have opened the edge $e$ if it was closed in $\omega$. 
If we let $\{ x \cc y \}$ denote the event that the end-points $x,y$ of $e$ are connected (possibly using the ghost), then
\begin{align*}
\kappa(\omega^e) = \kappa(\omega) \id_{\{x \cc y \} }(\omega)  +  (\kappa(\omega) -1) \id_{\{x \not \cc y \} }(\omega). 
\end{align*}
Thus, we get that $Y$ is increasing if and only if
\begin{align*}
Y(\omega_{\win}) \leq Y(\omega_{\win}^e)  = \frac{r_1^{o(\omega_{\win}^e)}\sum_{\omega_{\mathfrak{g}}} H_1^{o(\omega_{\mathfrak{g}})} \left(  q_1^{\kappa(\omega)} \id_{\{x\cc y \}}(\omega)  + \frac{1}{q_1} q_1^{\kappa(\omega)} \id_{\{x \not \cc y \}}(\omega) \right) }{r_2^{o(\omega_{\win}^e)}\sum_{\omega_{\mathfrak{g}}} H_2^{o(\omega_{\mathfrak{g}})} \left(  q_2^{\kappa(\omega)} \id_{\{x \cc y \}}(\omega)  + \frac{1}{q_2} q_2^{\kappa(\omega)} \id_{\{x \not \cc y \}}(\omega) \right)}. 
\end{align*}
Notice that for $i \in \{1,2 \}$ any event $A$ and configuration of internal edges $\omega_{\win},$ then
	\begin{align*}
	\phi_{p_i, q_i, h_i,G}[A \mid \omega_{\win}] = \frac{	\phi_{p_i, q_i, h_i,G}[A \cap \{ \omega_{\win} \}]}{	\phi_{p_i, q_i, h_i,G}[ \{ \omega_{\win} \}] } = \frac{r_i^{o(\omega_{\win})} \sum_{\omega_{\mathfrak{g}}} H_i^{o(\omega_{\mathfrak{g}})} q_i^{\kappa(\omega)} \id_A}{ r_i^{o(\omega_{\win})}\sum_{\omega_{\mathfrak{g}}} H_i^{o(\omega_{\mathfrak{g}})} q_i^{\kappa(\omega)}} = 	\frac{ \sum_{\omega_{\mathfrak{g}}} H_i^{o(\omega_{\mathfrak{g}})} q_i^{\kappa(\omega)} \id_A}{ \sum_{\omega_{\mathfrak{g}}} H_i^{o(\omega_{\mathfrak{g}})} q_i^{\kappa(\omega)}},
	\end{align*}
which makes the inequality above equivalent to 
\begin{align} \label{central_ineq} 
r_2 \left( \left(1- \frac{1}{q_2} \right)\phi_{p_2, q_2, h_2,G}[x \cc y \mid \omega_{\win}] + \frac{1}{q_2}  \right) \leq r_1 \left( \left(1- \frac{1}{q_1} \right)\phi_{p_1, q_1, h_1,G}[x \cc y \mid \omega_{\win}] + \frac{1}{q_1}  \right).
\end{align}
If $x$ and $y$ are connected with the inner edges then the condition is just $r_2 \leq r_1 $ which again is equivalent to $p_1 \geq p_2$. 
If we let $C_x$ be the cluster of $x$ in $\omega_{\win}$, then we can rewrite the previous inequality using  Lemma \ref{ghosted} to obtain
\begin{align*}
\frac{r_2}{q_2} \left( \left(q_2 -1 \right)\tanh_{q_2}(\abs{C_x}h_2)\tanh_{q_2}(\abs{C_y}h_2 ) + 1  \right) \leq  \frac{r_1}{q_1} \left( \left(q_1 -1 \right)\tanh_{q_1}(\abs{C_x}h_1)\tanh_{q_1}(\abs{C_y}h_1 ) + 1  \right). 
\end{align*}
The main statement now follows since $\abs{C_x}$ and $\abs{C_y}$ are positive integers.
 \end{proof}

\subsection{Upper bound on the Kert{\'{e}}sz line}
For any dimension $d \geq 2$ the random-cluster model has a phase transition for $q \in \lbrack 1, \infty) $ at some $p_c(q)$.
For example it is proven \cite{BD12} for $d=2$ that $p_c(q,0) = \frac{ \sqrt{q} }{1+ \sqrt{q}}$. It is proven in \cite{grimmett1995comparison} that $p_c(q)$ is strictly monotone and Lipschitz continuous in $q$ as a function from $\lbrack 1, \infty) \to \lbrack p_B,1)$ for all $d \geq 2$.\footnote{One may see that $p_c(q)$ is surjective as follows: For any $p$, the probability of crossing the annulus $\Lambda_{3}\setminus \Lambda_1$ can be made arbitrarily small by increasing $q$. The methods from Section 4.3 may then be employed to see that there is no percolation at $(p,q)$.} Therefore, it has an inverse function $q_c:\lbrack p_B,1) \to \lbrack 1, \infty)$ such that $(p, q_c(p))$ is critical. 
When $d= 2,$ inverting the relation before yields  $q_c(p,0) =  \left(\frac{p}{1-p}  \right)^2$. \\
\textit{Proof of Theorem \ref{thm:upper_bound}.} 
We know that for every $\varepsilon > 0 $ then there is percolation at $(p + \varepsilon,q_c(p), 0)$. Then, by Theorem \ref{eksplicitdom} we get that there is also percolation for $(p + \varepsilon, q,h)$ as long as 
\begin{align*}
\frac{1}{q_c(p,0)} \leq \frac{1}{q} \left( (q-1) \tanh_q(h)^2 + 1 \right). 
\end{align*} 
Then, for fixed $q$, since $\tanh_q$ is increasing, the minimal $h$ where the inequality is satisfied is the $h$ that satisfies the equality
$\sqrt{\frac{1}{q-1} \left(\frac{q}{q_c(p,0)} -1 \right)} = \tanh_q(h). $
Using again that $\tanh_q$ is strictly monotone yields the theorem. 
\qed\\

\begin{figure}
\captionsetup{width=.6\linewidth}
{  \caption{The upper bound for the Kert{\'{e}}sz line in $d=2$ proven in Theorem \ref{thm:upper_bound} (solid)  as well as Theorem \ref{thm:upper_bound_Ber} (dashed) for $q=1.1, 2, 10$ in red, blue and green respectively. The solid upper bound tends to $p_c(q,0)$ in the limit $h \to 0$. All the bounds have the correct limit as $h \to \infty$.}  \label{upper_bound}}{
     \includegraphics[scale =0.4]{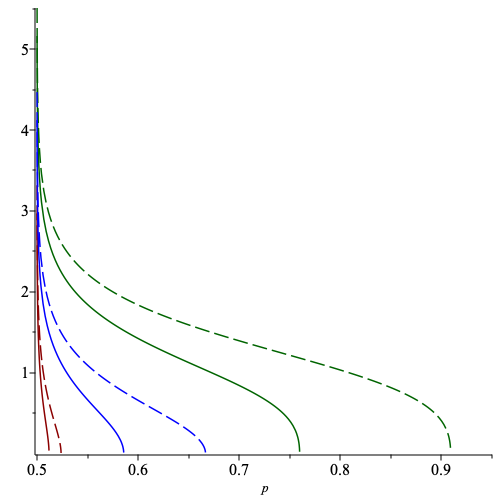}}
\end{figure}

Analytically, we can see that the bound has the correct behaviour in the limits. The bound becomes infinite whenever  $\frac{1}{q-1} \left(\frac{q}{q_c(p,0)} -1 \right) = 1$, i.e.\ when $q_c(p,0) =1,$  which again means that $p = p_B$.  Similarly, the bound tends to $0$ whenever $\frac{1}{q-1} \left(\frac{q}{q_c(p,0)} -1 \right) = 0$, i.e.\ $q = q_c(p,0),$ which means that $p = p_c(q,0)$. 
In Figure \ref{upper_bound} we have plotted this upper bound. Notice that, in contrast to the bounds from \cite{Rui}, it gives the correct limit when $h \to 0$.

\begin{remark}
One may note that our techniques allow one to use something about the random-cluster model for $q \neq 2$ to gain information about the Ising model where $q=2$. 
\end{remark}

\subsubsection*{Upper bound using the Bernoulli percolation threshold}
Before, we used knowledge about the phase transition at $h=0$ to infer knowledge about the  Kert{\'{e}}sz line at all $h \geq 0$. In the following, we use a similar trick where we use that if $p = p_B + \varepsilon $ where $p_B(d)$ is the critical $p$ of Bernoulli percolation, then there is percolation without using the ghost for $h$ sufficiently large. In $d=2$ we can then use Kesten's  celebrated result that $p_B(2) = \frac{1}{2}$ (see \cite{Kesten1980TheCP}). For $d=2$ the method does not produce a better bound than the bound in Theorem \ref{thm:upper_bound}, but if one only has knowledge about the phase transition of Bernoulli percolation in higher dimensions and not the random-cluster model, it can produce a better bound than Theorem \ref{thm:upper_bound}. 

\begin{theorem} \label{thm:upper_bound_Ber}
Let $q \in \lbrack 1, \infty)$ and $r_B = \frac{p_B}{1-p_B}$ where  $p_B(d)$ is the critical parameter of Bernoulli percolation in $d$ dimensions. Then, the following upper bound on the Kert{\'{e}}sz line holds: 
\begin{align*}
h_c(p) \leq \operatorname{arctanh}_{q} \left( \sqrt{ \frac{1}{q-1} \left( q r_B \frac{1-p}{p} -1 \right)} \right). 
\end{align*}
In particular, if $d=2$ and $q=2$ corresponding to the planar Ising model we obtain
\begin{align*}
h_c(p) \leq \operatorname{arctanh} \left( \sqrt{2 \frac{1-p}{p} - 1 } \right). 
\end{align*}  
\end{theorem} 
\begin{proof} 
In Theorem \ref{eksplicitdom} we let $(p_1,q_1,h_1) =(p_1,q,h_1) $ for some $p_1 \geq p_B$ as well as $(p_2,q_2,h_2) = (p_B + \varepsilon, q, N) $ for some small $\varepsilon > 0$ and arbitrarily large $N$. Then the condition for stochastic domination is that 
\begin{align} \label{suf_dom} 
r_2 \left( \left(q-1 \right) \tanh_{q}(h_2 n ) \tanh_{q}(h_2 m ) + 1  \right) \leq r_1 \left( \left(q-1 \right)\tanh_{q}(h_1 n ) \tanh_{q}(h_1 m )+ 1\right)
\end{align}
for all positive integers $n,m$. 
Notice that we have $\tanh_q(h_2 n) \leq 1 $ and therefore 
\begin{align*}
r_2 \left( \left(q-1 \right) \tanh_{q}(h_2 n ) \tanh_{q}(h_2 m ) + 1  \right) \leq q r_2. 
\end{align*}
Since $n,m \geq 1$ then $\tanh_q(h_1 n ) \geq \tanh_q(h_1) $ and thus 
\begin{align*}
 r_1 \left( \left(q-1 \right)\tanh_{q}(h_1)^2+ 1\right) \leq r_1 \left( \left(q-1 \right)\tanh_{q}(h_1 n) \tanh_{q}(h_1 m)+ 1\right).
\end{align*}
Thus, it is sufficient for (\ref{suf_dom}) and therefore stochastic domination that 
\begin{align*}
q r_2 \leq r_1 \left( \left(q-1 \right)\tanh_{q}(h_1)^2 + 1\right). 
\end{align*} 
Picking $r_2 = r_B = \frac{p_B}{1- p_B}$ as well as recalling that $r_1 = \frac{p}{1-p}$ we obtain that it is sufficient that 
\begin{align*}
\arctanh_{q} \left( \sqrt{ \frac{1}{q-1} \left( q r_B \frac{1-p}{p} -1 \right)} \right) \leq h_1. 
\end{align*} 
To finish the proof, we note that there is percolation at $(p_2,q_2,h_2)$ so long as $N$ is chosen large enough.
\end{proof}

\subsection{Lower bound on the Kertész line}
In this section, we give a lower bound on the Kertész line following the strategy in \cite{Renormalization} for proving exponential decay of cluster sizes (which, in particular, implies a lack of percolation). The arguments rely on nothing but sharpness of the subcritical phase (as is known from \cite{sharpness}). 
\begin{lemma} \label{separation lemma}
Let $S$ be a finite subset of $\mathbb{Z}^d$. Then, there exists a subset $T_S\subseteq S$ such that $|T_S|\geq \frac{|S|}{4^d}$ and for every $v,w\in T_S,$ we have that $\|v-w\|_{L^1} \geq 4.$
\end{lemma}

\begin{remark} By considering $S=\Lambda_k$ and letting $k\to \infty,$ we get that the bound is sharp.
\end{remark}
\begin{proof}
Note that 
$$
S= \bigcup_{\tau\in [0,3]^d} S\cap (\tau+4\mathbb{Z}^d),
$$
implying that there exists $\tau_0\in [0,3]^d$ such that $|S\cap (\tau_0+4\mathbb{Z}^d)|\geq \frac{|S|}{4^d}$. It is easily seen that $T_S:=S\cap (\tau_0+4\mathbb{Z}^d)$ also has the desired separation property.
\end{proof}
In the following, we let $A_n$ denote the set of connected subsets of $\mathbb{Z}^d$ containing $0$ and exactly $n$ vertices. It is a classic result that $\abs{A_n}$ has an exponential growth rate. For instance, Lemma 5.1 in \cite{Kesten82} shows that $\abs{A_n}\leq \left(\frac{(2d+1)^{2d+1}}{(2d)^{2d}}\right)^n
:=\mu^n$.\\

\textit{Proof of Theorem \ref{thm:lower_bound}.}
For each $v \in 2k \Z^d,$ define $X(v)  = \id \lbrack \Lambda_k(v) \cc \partial \Lambda_{3k}(v)  \rbrack,$  which is a site percolation process on $2k \Z^d$. 

Given a $v \in 2k \Z^d,$ we let $C_\omega(v)$ be the cluster of $v$ in $\omega_{in}$, $C_X(v)$ be the set of $w\in 2k\Z^d$ such that $d(w,C_{\omega}(v))\leq k$. Note that if $|C_{\omega}(v)|>3^d|\Lambda_k|,$ then every $w\in C_X(v)$ is open in $X$. 
Furthermore, if $\abs{C_\omega(v)} \geq N$ then $ \abs{C_X(v)} \geq  \left \lfloor \frac{N}{|\Lambda_{k}|}  \right \rfloor := n$.

To account for the magnetic field, we will need to control the density of ghost edges. For each $v \in 2k \Z^d,$ we say that $v$ is \emph{good} if every ghost edge in $\Lambda_{3k}(v)$ is closed. Otherwise, we say that $v$ is \emph{bad}. Interchangeably, we will say that the box itself is good respectively bad. We denote the process of bad boxes by $\mathcal{B},$ i.e.\ $\mathcal{B}(v)=\id[v\textrm{ is bad}]$.

We will now let $N>3^d|\Lambda_k|$ and bound the quantity $\phi_{p,q,h}( \abs{C_\omega(v)} \geq N)  \leq \phi_{p,q,h}( \abs{C_X(v)} \geq n) $. 
If $\abs{C_X(v)} \geq n,$ then we know that there is an open connected $S$ in $X$ containing $v$ and exactly  $n$ vertices. 
Hence, by a union bound,
$$
 \phi_{p,q,h,\mathbb{G}}[\abs{C_X(v)} \geq n]   \leq \sum_{\substack{S\subseteq 2k\mathbb{Z}^d, \\
 S/(2k)\in A_n}} \phi_{p,q,h,\mathbb{G}}[S\textrm{ open in } X]. 
$$
Now, by Lemma \ref{separation lemma}, we can pick a thinned set $T_S$ of at least $\frac{|S|}{4^d}$ vertices, such that for $w,w'\in T_S$, we have $\Lambda_{3k}(w)\cap \Lambda_{3k}(w')=\{ \mathfrak{g} \}$ \footnote{Notice that, since $\Lambda_{3k}(w)$ and $\Lambda_{3k}(w')$ only have a single vertex in common, the wired measure on the union is a product measure of the wired measure on each box.}. For a given $w\in T_S$, we have, by the Domain Markov Property, that if $E^c_k(w)$ denotes the complement of the edges between vertices in $\Lambda_{3k}(w)$,
$$
\phi_{p,q,h,\mathbb{G}}[ X(w)=1 | \; \omega|_{E^c_k(w)}]\leq \phi_{p,q,h,\mathbb{G}}[\mathcal{B}(w)=1|\; \omega|_{E^c_k(w)}]+\phi_{p,q,0,\Lambda_{3k}(w)}^1[X(w)=1].
$$
Furthermore, by Theorem \ref{comparison theorem},  we have
$$
\phi_{p,q,h,\mathbb{G}}[\mathcal{B}(w)=1|\; \omega|_{E^c_k(w)}]\leq 1-(1-p_h)^{|\Lambda_{3k}|}.
$$
Consequently,
$$
\phi_{p,q,h,\mathbb{G}}\left[\bigcap_{w\in T_S} \{X(w)=1\}\right]\leq \prod_{w\in T_S} \left(\phi_{p,q,0,\Lambda_{3k}(w)}^1[X(w)=1]+1-(1-p_h)^{|\Lambda_{3k}|} \right).
$$

 \begin{figure}
\captionsetup{width=.6\linewidth}
{  \caption{A representation of the coarse-graining scheme in the proof of Theorem \ref{thm:lower_bound}. The dots illustrate the grid $2k \Z^2$. With the dotted lines we indicate the boxes $\Lambda_{k}(v)$ and $\Lambda_{3k}(v)$ around a point $v$. Notice that since the cluster $S$ enters the box $\Lambda_{k}(v)$ this means that $X(v) =1$. With the red dots we indicate bad sites in $2 k \Z^2$. In the proof, we use that, for sufficiently small $h$, these bad sites are very uncommon.  \label{lower_bound_sketch}}}{
     \includegraphics[scale =0.3]{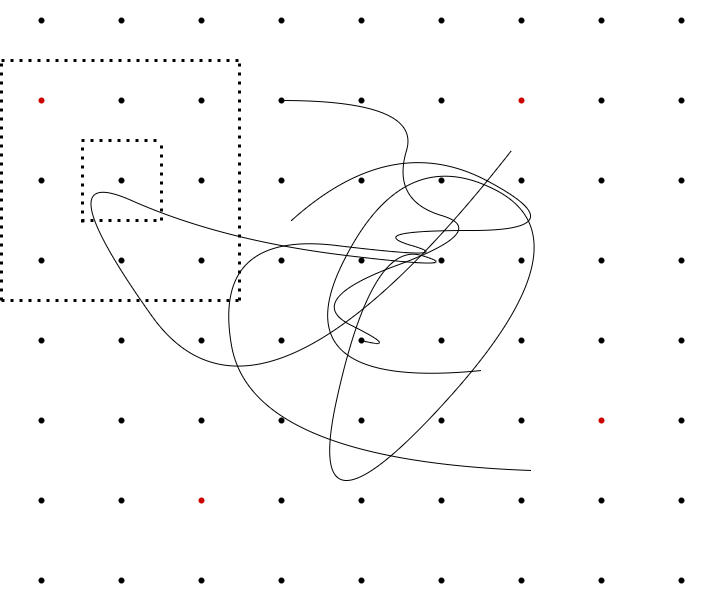}}
\end{figure}

\noindent Adding all of this together yields
\begin{align*}
 \phi_{p,q,h,\mathbb{G}}[ \abs{C_X(v)} \geq n]   &\leq \sum_{\substack{S\subseteq 2k\mathbb{Z}^d,\\
 S/(2k)\in A_n}} \phi_{p,q,h,\mathbb{G}}[S\textrm{ open in } X]  \\
 &\leq \sum_{\substack{S\subseteq 2k\mathbb{Z}^d,\\
 S/(2k)\in A_n}} \phi_{p,q,h,\mathbb{G}}[\cap_{w\in T_S} \{X(w)=1\}]\\
 &\leq \sum_{\substack{S\subseteq 2k\mathbb{Z}^d,\\
 S/(2k)\in A_n}} \prod_{w\in T_S} \left( \phi_{p,0,\Lambda_{3k}(w)}^1[X(w)=1]+1-(1-p_h)^{|\Lambda_{3k}|}\right)\\
 &\leq \sum_{\substack{S\subseteq 2k\mathbb{Z}^d,\\
 S/(2k)\in A_n}} (\phi_{p,q,0,\Lambda_{3k}(w)}^1[X(w)=1]+1-(1-p_h)^{|\Lambda_{3k}|})^{|S|/4^d}\\
 &\leq \left( \mu^{4^d} \left(\phi_{p,q,0,\Lambda_{3k}(0)}^1[X(0)=1]+1-(1-p_h)^{|\Lambda_{3k}|}\right) \right)^{n/4^d},
\end{align*}
which decays to $0$ if $1-(1-p_h)^{|\Lambda_{3k}|}<\frac{\delta}{2}$. This is attained for
$$ 
p_h<1-\left(1- \frac{1}{2\mu^{4^d}}\right)^{1/|\Lambda_{3k}|}=1-\left(1-\frac{\delta}{2}\right)^{1/|\Lambda_{3k}|}.
$$
\qed

We conclude this section with a discussion of the bounds and the tangent  at $(p_c,q,0)$ .
\subsection{Discussion of the tangent at \texorpdfstring{$(p_c,q,0)$ for $d=2$}{criticality}}
We now use the correlation length to get a slightly more explicit lower bound under the conjectural assumption of the existence of critical exponents. In the following we consider fixed $q \in \lbrack 1, 4 \rbrack$.  As in \cite[(1.5)]{duminilcopin2020planar} we define the correlation length $\xi$ for $h=0$ for $p < p_c$ by
\begin{align*}
    \xi = - \lim_{n \to \infty} \frac{n}{\log \left( \phi^1_{p,q,0} \left( 0 \cc  \partial \Lambda_n \right)  \right)} . 
\end{align*}
By a union bound, we have
$$
\phi^1_{p,q,0} \left[ 0 \cc  \partial \Lambda_{3n} \right]\leq \phi^1_{p,q,0, \Lambda_{3n} } \left[ \Lambda_n \cc  \partial \Lambda_{3n} \right]\leq \phi^1_{p,q,0, \Lambda_{3n}} \left[ \cup_{x\in \partial_v \Lambda_{n}} ( x \cc  \partial \Lambda_{3n} ) \right]\leq 8n\phi^1_{p,q,0, \Lambda_{3n}}[0 \cc \partial \Lambda_{2n}]
$$
implying that
$$
e^{-\frac{3n}{\xi}+o(n)} \leq \phi^1_{p,q,0, \Lambda_{3n}} \left( \Lambda_n \cc  \partial \Lambda_{3n} \right)\leq e^{-\frac{2n}{\xi}+o(n)}. 
$$
It is further expected that there exist constants $C_1,C_2, C >0$ such that
\begin{align} \label{eq:expected}
C_1 e^{-C\frac{n}{\xi}} \leq \phi^1_{p,q,0, \Lambda_{3n}} \left( \Lambda_n \cc  \partial \Lambda_{3n} \right)\leq C_2 e^{-C\frac{n}{\xi}} . 
\end{align}
uniformly in $p$ for fixed $q$. In the following, we will assume \eqref{eq:expected}. 
Then,
$$
C_2 e^{- C \frac{n}{\xi}}< \frac{\mu^{-16}}{2}
$$
is satisfied for 
$$
 n>  \frac{16}{C} \xi \log(\mu)  + C_3. 
$$
So choose $k =  \frac{16}{C} \xi \log(\mu)  + C_3 +1$. 
Now, by Bernoulli's inequality, we see that  
$$ 
\frac{1}{18 \mu^{16} k^2} <1-\left(1- \frac{1}{2\mu^{16}}\right)^{1/|\Lambda_{3k}|}.   
$$
Thus, it is sufficient for the assumption in Theorem \ref{thm:lower_bound} that
\begin{align*}
p_h < \frac{1}{18 \mu^{16} k^2} =  \frac{1}{18\mu^{16}(\frac{16}{C} \xi \log(\mu)  + C_3+1)^2 } = \frac{1}{C_4 \xi^2 + O(\xi)}. 
\end{align*}
in the limit $\xi \to \infty$. 
Now, in the planar case $d=2$, the correlation length $ \xi(p)$ is conjectured \cite{duminilcopin2020planar} to have the form 
\begin{align*}
\xi(p) \sim C \abs{p-p_c}^{- \nu} 
\end{align*}
for some critical $q$ dependent exponent $\nu$ which has the form 
\begin{align*}
    \nu(q)= \frac{2\arccos \left( - \frac{ \sqrt{q}}{2} \right) }{6\arccos \left( - \frac{ \sqrt{q}}{2} \right)  - 3 \pi} \in \left  \lbrack \frac{2}{3}, \frac{4}{3} \right \rbrack 
\end{align*}
for $q \in \lbrack 1, 4 \rbrack$. In the particular case of the Ising model, the conjecture is that $\nu(2) = 1$. 
Thus, under that conjecture using $\mu = \frac{5^5}{4^4}$, our condition becomes
\begin{align*}
p_h < \frac{1}{18\mu^{16} k^2} =  C \abs{p-p_c}^{2\nu}
\end{align*}
for some constant $C>0$. Using the fact that $p_h(h) = 1- e^{-2h}$ is approximately linear in $h$ for small $h,$ we see that the tangent of the bound is asymptotically flat as $ p \to p_c$.  The bound has horizontal tangent and this holds for all $1 \leq q \leq 4$ since it is conjectured that $2 \nu(q) > 1$ for all $q \in \lbrack 1, 4 \rbrack$. 
\\
\\
In conclusion, we notice that both our upper and lower bounds tend to the correct value $(p_c,q,0)$ as $h \to 0$. Thereby, the bounds complement those of \cite{Rui}, that are best for large $h$, i.e.\ around the Bernoulli percolation threshold. However, in our bounds the asymptote for the lower bound is horizontal and for the upper bound, it is vertical as shown in Figure \ref{upper_bound}. Since the lower bound is asymptotically horizontal, this leaves open the natural question of what the inclination of the tangent is at the point  $(p_c,q,0)$.\\
 \; Numerical evidence for the planar Ising case \cite{FS01} observed $\frac{\beta}{\beta_c} -1 \sim c h^\kappa $ in the limit $h \to 0$ for a $\kappa = 0.534(3)$, where it is noticed that $\frac{1}{\beta \delta} = \frac{8}{15} \approx 0.533$ (for the definition of the critical exponents $\beta, \delta$ see \cite{FS01}). Now, using that $\beta - \beta_c$ and $p-p_c$ have a linear relationship when both are small yields that $p-p_c \sim c h^\kappa$ or equivalently that 
\begin{conjecture}
In the limit $p \to p_c$ it holds for some constant $c >0$ that
\begin{align*}
    h_c(p) \sim c (p-p_c)^{\frac{15}{8}}. 
\end{align*}
\end{conjecture}

This indicates that the lower bound is almost optimal and one would have to improve the upper bound.
In, particular this leaves plenty of further work concerning the asymptotic behaviour of the  Kert{\'{e}}sz line for $h \to 0$. A potential starting point for this program could be \cite[Lemma 8.5]{duminilcopin2020planar} which was also used in \cite{klausen2022mass} to gain knowledge about the correlation length in a non-zero magnetic field. 

\section{Continuity of the Kert{\'{e}}sz line phase transition} \label{continuity} 
There are several ways to define continuity of a phase transition: One relates to whether or not $\theta(p_c)=0$ (or, equivalently, $\mu^1_{\beta_c}(\sigma_0 \cdot \boldsymbol{1})=0$)  and another to whether the infinite-volume measures $\phi^1_{p_c}$ and $\phi^0_{p_c}$ coincide or not (or whether $\mu^1_{\beta_c}=\mu^0_{\beta_c}$). A third perspective pertains to the regularity of the pressure
$$
\mathfrak{Z}=\lim_{n\to\infty} \frac{\log(Z^1(\Lambda_n))}{|\Lambda_n|},
$$
where $Z^1$ denotes the partition function of either the random-cluster model or the Potts model. \\
Indeed, one can show that $\mathfrak{Z}$ is convex as a function of $p$ (or $\beta$) and hence, admits left and right derivatives everywhere (see \cite[Exercise 21]{DC17}). These take the form 
\begin{align*}
\frac{\partial}{\partial p^+} \mathfrak{Z}=\phi^1_{p,q,h}[\omega_e] \hspace{1cm} \frac{\partial}{\partial p^-} \mathfrak{Z}=\phi^0_{p,q,h}[\omega_e],
\end{align*}
where $\phi^b_{p,q,h}[\omega_e]$ denotes the probability that a given inner edge $e$ is open \footnote{By translation invariance of the infinite-volume measure, this probability does not depend on the choice of inner edge.}. \\
Furthermore, by Proposition 4.6 in \cite{Gri06}, $\phi^1_{p,q,h,\mathbb{Z}^d}=\phi^0_{p,q,h,\mathbb{Z}^d}$ if and only if these probabilities agree. Accordingly, these two measures are different at $p_c$ if and only if $\mathfrak{Z}$ fails to be $C^1$. Furthermore, non-uniqueness of the infinite volume measure would imply that $\theta(p_c)>0$. To see this last implication, one might simply note that for any increasing event $A$ depending only on finitely many edges,
$$
\phi^1_{p,q,h,\mathbb{Z}^d}[A\cap (0\not\cc \infty)]=\sup_{k} \phi^1_{p,q,h,\mathbb{Z}^d}[A\cap (0\not\cc \Lambda_k)]\leq \sup_{k} \phi_{p,q,h,\Lambda_k}^0[A]=\phi_{p,q,h,\mathbb{Z}^d}^0[A]
$$
and hence, $\phi^1_{p,q,h}=\phi^0_{p,q,h}$ if $\phi^1_{p,q,h}[0\cc \infty]=0$. 
In particular, this means that a discontinuous geometric phase transition implies a (first order) thermodynamic one. By contraposition, this means that the analyticity of the pressure, which we prove below, rules out only thermodynamic phase transitions. 

The converse direction, that $\theta(p_c)>0$ if and only $\phi^1_{p_c}\neq \phi^0_{p_c},$ is subtle, since a general proof would imply that $\theta(p_c,1,0)=0,$ which remains perhaps the single largest open question in all of percolation theory. Indeed, in \cite{Books}, an example is given of several random-cluster models which have unique infinite volume measures at $p_c$, but such that $\theta(p_c)>0$, meaning that the converse being true would have to rely upon the specific structure of $\mathbb{Z}^d$.
\\
Thus, we shall spend this section focusing on the characterisation of phase transition via regularity of the pressure $\mathfrak{Z}$. Following \cite{velenik}, we extract a cluster expansion for the random-cluster model in order to prove Theorem \ref{analytic pressure}.

For the proof, we start with the case of the Potts model.
\begin{lemma}
For any finite subgraph $G$ of an infinite graph $\mathbb{G}$ and $\beta,h>0$, we have
\begin{align*}
Z^{1,q}_{\beta,h}(G) =& e^{\beta |E|} e^{h|V|}\sum_{G'=(V',E') \subseteq G}  e^{-\beta |E'|}\tilde{Z}^{0,q-1}_{\beta,0}(G') e^{-(1+\frac{1}{q-1})(h|V'|+\beta|\partial_e G'|)}\\
:=& e^{\beta|E|+h|V|} \Xi^{q}_{\beta,h}(G),
\end{align*}
where $\tilde{Z}^{0,q-1}$ is constructed by taking spins amongst the $q-1$ spins in $\mathbb{T}_q$ different from $\mathbf{1}$.
\end{lemma}
\begin{proof}
For $\sigma\in \mathbb{T}_q^V$, let $G'(\sigma)$ denote the subgraph of $G$ with vertex-set $\{i\in V|\; \sigma_i\neq \mathbf{1}\}$. Then, clearly
\begin{align*}
Z^{1,q}_{\beta,h}(G) &=\sum_{G'=(V',E')\subseteq G} \sum_{\substack{\sigma\in \mathbb{T}_q\\ G'(\sigma)=G'}} e^{-\beta\mathcal{H}^1(\sigma)+h\sum_{i\in V} \langle \sigma_i,\mathbf{1}\rangle} \\
&=\sum_{G'=(V',E')\subseteq G} e^{\beta |E\setminus (E'\cup \partial_e G')|} e^{h| V\setminus V'|} e^{-\frac{\beta}{q-1} |\partial_e G'|} e^{-\frac{h}{q-1} \vert V' \vert}\sum_{\sigma\in (\mathbb{T}_q\setminus \{\mathbf{1}\})^{V'}}e^{-\beta H^0(\sigma)} \\
&=e^{\beta|E|} e^{h |V|}\sum_{G'=(V',E')\subseteq G} e^{-h(1+\frac{1}{q-1})|V'|}e^{-\beta(1+\frac{1}{q-1})|\partial_e G'|} e^{-|E'|}\tilde{Z}^{0,q-1}_{\beta,0},
\end{align*} 
which is what we wanted.
\end{proof}
If $H$ and $H'$ are two finite, disjoint subgraphs of a larger graph $G$, then clearly, $\tilde{Z}^{b,q}_{\beta,h}(H\cup H')=\tilde{Z}^{b,q}_{\beta,h}(H) \tilde{Z}^{b,q}_{\beta,h}(H')$. Hence, by decomposing $G'$ into a tuple of connected components $\mathcal{S}=(S_1,S_2,...,S_m)$, we get that
$$
\Xi^q_{\beta,h}(G)=\sum_{\mathcal{S}\in \Gamma} \prod_{S\in \mathcal{S}} w^q_{h,\beta}(S) \prod_{S\neq S'\in \mathcal{S}} \delta(S,S'),
$$
where $\Gamma$ is the set of connected subgraphs of $G$ and for $S=(V_S,E_S)$,
$$
w_{\beta,h}^q(S)= e^{-\beta|E_S|}(S)e^{-(1+\frac{1}{q-1})(h|V_S|+\beta|\partial_e S|)}\tilde{Z}^{0,q-1}_{\beta,0}
$$
and 
$$
\delta(S,S')=\id_{\{d(S,S')>1\}},
$$
where $d(S,S')=\inf_{x\in S,y\in S'}\| x-y\|_{\ell^1}$.
Thus, we have written $\Xi^q_{\beta,h}(G)$ as the partition function of a polymer model with hardcore interactions. \\
It is worth noting that, since $q\geq 2$,
$$
\tilde{Z}^{0,q-1}_{\beta,0}(S)\leq (q-1)^{|V_S|} e^{\beta |E_S|},
$$
so that, in fact,
$$
w_{\beta,h}^q(S)\leq (q-1)^{|V_S|}e^{-(1+\frac{1}{q-1})(h|V_S|+\beta|\partial_e S|)}.
$$
\begin{lemma}
Assume that $G\subseteq \mathbb{Z}^d$, let $B(S)$ denote the set of vertices within $\ell^1$ distance $1$ of $S$ and $a(S)=|B(s)|$. Then, if $\Gamma^{\infty,d}$ is the set of all connected subgraphs of $\mathbb{Z}^d$ and $S'$ denotes some fixed element of $\Gamma$, we get that
$$
\sum_{S\in \Gamma} w_{\beta,h}^q(S) e^{a(S)} (1-\delta(S,S'))\leq a(S') \max_{j\in B(S')} \sum_{j\in S\in \Gamma^{\infty,d}} w_{\beta,h}^q(S) e^{a(S)}.
$$
\end{lemma}
\begin{proof} Note that the only positive contributions to the sum on the left-hand side come from $S$ such that $d(S,S')\geq 1.$ Thus,
$$
\sum_{S\in \Gamma} w_{\beta,h}^q(S) e^{a(S)} (1-\delta(S,S'))\leq \sum_{j\in B(S')} \sum_{j\in S\in \Gamma} w_{\beta,h}^q(S) e^{a(S)},
$$
from which the lemma follows immediately.
\end{proof}
Recall the function $h_0$ from Theorem \ref{analytic pressure}.
\begin{lemma}
For any $h>h_0(q,d),$ we have
$$ \sum_{0\in S\in \Gamma^{\infty,d}} w_{\beta,h}^q(S) e^{a(S)}<1.$$
\end{lemma}
\begin{remark}
By translation invariance, it suffices to consider $j=0$.
\end{remark}
\begin{proof}
We see that
\begin{align*}
\sum_{0\in S\in \Gamma^{\infty,d}} w_{\beta,h}^q(S) e^{a(S)} &=\sum_{k=1}^{\infty} e^{-(1+\frac{1}{q-1})hk}\sum_{\substack{0\in S\in \Gamma^{\infty,d}\\ |V_S|=k}} e^{(a(S))} e^{-\beta|E_S|} \tilde{Z}_{\beta,0}^{0,q-1}(S)e^{-(1+\frac{1}{q-1})\beta |\partial_e S|} \\
&\leq \sum_{k=1}^{\infty} (q-1)^k e^{-(1+\frac{1}{q-1})hk} e^{2d k} |\{S\in \Gamma^{\infty,d}|\; |V_S|=k, \; 0\in V_S \}|\\
&\leq \sum_{k=1}^{\infty} (q-1)^k e^{-(1+\frac{1}{q-1})hk} e^{2d k} \left(\frac{(2d+1)^{2d+1}}{(2d)^{2d}}\right)^k,
\end{align*}
where, in the last line, we have once again used  Lemma 5.1 in \cite{Kesten82}.
The right hand side is, of course, a geometric sum which is convergent with a sum less than $1$ for $h>h_0(q,d)$.
Thus, this case of Theorem \ref{analytic pressure} follows from \cite[Theorem 5.4]{velenik}. 
\end{proof}

\begin{corollary}
The above results carry over to the partition function of the random-cluster model with wired boundary conditions for non-integer $q$.
\end{corollary}
\begin{proof}
For integer $q$, note that, by the Edwards-Sokal Coupling, we have
$$
\Xi^q_{\beta,h}(G)=e^{-\beta |E|}e^{-h |V|}Z^{1,q}_{\beta,h}(G)=\sum_{\omega_{in} \in \{0,1\}^E} \sum_{\omega_{\mathfrak{g}} \in \{0,1\}^V} p^{o(\omega_{in})}(1-p)^{c(\omega_{in})} p_h^{o(\omega_{\mathfrak{g}})}(1-p_h)^{c(\omega_{\mathfrak{g}})} q^{\kappa^1(\omega,\omega_\mathfrak{g})}
$$
for $\beta=-\frac{q-1}{q} \log(1-p)$ and $h=-\frac{q-1}{q}\log(1-p_h)$.\\*
Note that only the pairing of $\beta$ and $p$ depends on the inner product in $\mathbb{T}_q$, whereas the role of the factor of $q^{\kappa(\omega)}$ is to cancel with the uniformly random colouring in the Edwards-Sokal coupling. \\ Therefore, when considering fewer colours, we still have
$$
e^{-\beta |E_S|}\tilde{Z}^{0,q-1}_{\beta,0}(S)=\sum_{\omega\in \{0,1\}^{E_s}} p^{o(\omega)} (1-p)^{c(\omega)} (q-1)^{\kappa(\omega)},
$$
Hence, redefining 
$$
w_{p,h}^q(S)= \sum_{\omega\in \{0,1\}^{E_s}} p^{o(\omega)} (1-p)^{c(\omega)} (q-1)^{\kappa(\omega)}e^{-(1+\frac{1}{q-1})(h|V_S|+\beta(p)|\partial_e S|)}
$$
represents the random-cluster model partition function with wired boundary conditions as a polymer model for arbitrary $q\geq 1$. Here, we have inverted the formula (\ref{edge weights}) to get $\beta(p)=\frac{q-1}{q}\log(1-p)$. \\
For $q\geq 2$, since any vertex belongs to at most one component, it remains true that 
\begin{align*}
\sum_{\omega\in \{0,1\}^{E_s}} p^{o(\omega)} (1-p)^{c(\omega)} (q-1)^{\kappa(\omega)}\leq \sum_{\omega\in \{0,1\}^{E_s}} p^{o(\omega)} (1-p)^{c(\omega)} (q-1)^{|V_S|}=(q-1)^{|V_S|},
\end{align*}
and so, we retain the convergent cluster expansion above with the same bounds.\\
For $q\in (1,2),$ we instead use that $(q-1)^{\kappa(\omega)}\leq (q-1)$ to get
$$
\sum_{\omega\in \{0,1\}^{E_s}} p^{o(\omega)} (1-p)^{c(\omega)} (q-1)^{\kappa(\omega)}\leq q-1.
$$
Thus, in this case,
$$
\sum_{0\in S\in \Gamma^{\infty,d}} w^q_{p,h}(S)e^{a(S)}\leq (q-1)\sum_{k=1} e^{-(1+\frac{1}{q-1})hk}e^{2dk} \left(\frac{(2d+1)^{2d+1}}{(2d)^{2d}}\right),
$$
and we once again get convergence for $h>h_0(q,d)$.
\end{proof}

In \cite{Blanchard_2008}, discontinuity was proven using the Pirogov-Sinai theory.  There, it was proven  for $d \geq 2, q  \geq 1, h \geq 0$ that if 
\begin{align*}
c_d (1 + (q-1)e^{-h})^{-\frac{1}{2d}}<1   
\end{align*}
for some inexplicit constant $c_d,$ which only depends on the dimension, then the phase transition on the  Kert{\'{e}}sz line is discontinuous - i.e.\ $\mathfrak{Z}$ fails to be $C^1$. See also analogous results for the Potts model from \cite{bakchich1989phase}. 
To plot this in the $(q,h)$ plane, we can rearrange it into the condition that $h < \log \left( \frac{q-1}{c_d^{2d} -1} \right) $. 

 \begin{figure}
\captionsetup{width=.8\linewidth}
{ \caption{Plot of regions of discontinuity and continuity in the $(q,h)$ plane. For any $q\geq 1$ and $h\geq 0$ there is a unique $p(q,h)$ such that $(p(q,h),q,h)$ is critical. We colour the plane red or blue according to whether the phase transition at that point is a continuous or discontinuous. 
In Theorem \ref{analytic pressure} above we prove continuity of the phase transition in the contiguous region coloured red. Continuity on the line $q=2$ follows from the Lee-Yang theorem \cite{lee1952statistical} and for $h=0$ and $q \in \lbrack 1, 4 \rbrack$ from the explicit proof \cite{DuminilCopin2017ContinuityOT}. The blue region is not proven to be discontinuous since the constant $c_d$ is not explicit. Instead, the blue region is obtained from using the best possible constant in the bound from  \cite{Blanchard_2008} (this constant is $c_d =\sqrt{2}$ to match the change between continuous and discontinuous phase transition at $h=0$ for $q=4$).}  \label{continuity_sketch}}{
     \includegraphics[scale =0.4]{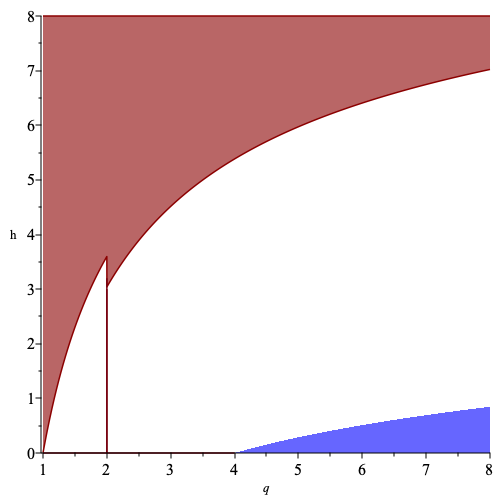}}
\end{figure}

\section{Outlook} \label{outlook}
We end by giving an outlook introducing further problems on Kert{\'{e}}sz line which lie in natural continuation of our work.  
\subsection*{Continuity of discontinuity}
For $h=0$, the quantity $\phi_{q,p_c(q)}^1( 0 \cc \infty) \searrow 0$ as $q \searrow 4 $. We conjecture that this is also the case for $h>0$: 
\begin{conjecture}
The set of $(q,h)$ such that the function $p\mapsto \mathfrak{Z}$ is not $C^1$ is open. 
\end{conjecture}

One could choose to view this as continuity of the gap 
\begin{align*}
\phi_{p_c(q,h),q,h,\mathbb{Z}^d}^1[0  \cc  \infty] - \phi_{p_c(q,h),q,h,\mathbb{Z}^d}^0[0  \cc  \infty].
\end{align*}
Continuity of the gap for $h=0$ was proven in \cite{DuminilCopin2017ContinuityOT, Diskont2016} by explicit computation of the critical magnetization which was continuous around $q=4$. 
One might note that in the special case of $d = 2$, the techniques of \cite{Renormalization} do apply more or less verbatim to the case where $h > 0$, implying that the regime of the exponential decay of the truncated wired measure is open in the $(p, h)$-plane. Seeing as this is a purely infinite volume phenomenon, however, makes attacking it quite delicate. 

Furthermore, we conjecture the following.
\begin{conjecture}
For each $q \in \lbrack 1, \infty)$ the map $h \mapsto \phi_{p_c(q,h),q,h,\mathbb{Z}^d}^1[0 \cc \infty]$ is decreasing. 
\end{conjecture}
Under this conjecture, we would see that the discontinuity in the $(q,h)$ plane (as plotted on Figure \ref{continuity_sketch}) would be a contiguous region. 
This would further establish the existence of a tri-critical point defined by 
\begin{align*}
h_c(q) = \inf \{ h \geq 0 \mid \phi_{p_c(q,h),q,h,\mathbb{Z}^d}^1[0 \cc \infty] = 0 \}. 
\end{align*}
In the physics literature \cite{KS}, there are some numerical studies which give rise to predictions of the universality class of this phase transition  which at this point in time seems out of reach.
One simple point that we can make along these lines is that if $1 \leq q \leq 4$ and $h \to 0$ then, by easy stochastic domination arguments, it holds that $\phi_{(p_c(q,h),q,h)}^1[0  \cc  \infty] \to 0$. The conjecture would imply that $\phi_{p_c(q,h),q,h,\mathbb{Z}^d}^1[0  \cc  \infty]$ is identically 0. 
 
\subsection*{Pseudo-critical line}
Since the Kert{\'{e}}sz line characterizes a geometric rather than a thermodynamic phase transition it is a priori not clear whether any signs of the phase transition transfer from the random-cluster model to the Potts model. Of course, in the regime of discontinuous phase transition (large $q$ and small $h$), the discontinuity implies that the free energy is not $C^1$ and therefore, the existence of a thermodynamic phase transition.

However, in the case where the  Kert{\'{e}}sz line  does not necessarily correspond to a thermodynamic phase transition we can ask whether any feature of the Kert{\'{e}}sz line can be observed in the Potts model. For general $q \in (1, \infty),$ one might, as in \cite{FS01}, define a \emph{pseudocritical line} as the line where the susceptibility is maximal in $p$.  Since divergent susceptibility is an indicator of criticality, it is a natural question to ask whether the psedocritical line and the Kert{\'{e}}sz line coincide, i.e.\ whether the susceptibility is maximal exactly on the Kert{\'{e}}sz line or not. If it is not the case, as it is conjectured in \cite{FS01}, it would be interesting to investigate whether, at $q>4$, the susceptibility peaks at a different value than the critical line and why. 
Similarly, one might consider the line of the maximal correlation length and ask to what extent that line coincides with the  Kert{\'{e}}sz line. 

Finally, one may ask about the critical exponents of the Kert{\'{e}}sz line. For example, it is claimed in \cite{FS01} that in the planar Ising case $q=d=2$, the critical exponents correspond to Bernoulli percolation rather than those of the FK-Ising.

\subsection*{Kert{\'{e}}sz line for the random current and loop $\mathrm{O}(1)$ models}
It is also natural to consider the  Kert{\'{e}}sz line problem for the random current and loop $\mathrm{O}(1)$ models. 
Let $P^\emptyset_h$ denote the random current measure and let $P^{\otimes 2}_h$ denote the sourceless double random current (see \cite{DC17} for definitions of the measures). Just as we have done in this article, a magnetic field can be implemented with a ghost vertex, allowing us to once again consider the percolation phase transition. However, these models lack monotonicity \cite{klausen2021monotonicity}, making the problem much more intricate. However, the monotonicity required to establish the existence of the  Kert{\'{e}}sz line  is much weaker than the overall monotonicity where the counterexamples of \cite{klausen2021monotonicity} apply. This leads us to the following conjectures that would establish the existence of the Kert{\'{e}}sz line for random currents. 
\begin{conjecture}
For any $d \geq 2$ the functions $
h \mapsto P^\emptyset_h \left[0 \overset{\Z^d}{\cc}  \infty \right] $ and $
h \mapsto P^{\otimes 2}_h \left[] 0 \overset{\Z^d}{\cc}  \infty \right] $  are increasing. 
\end{conjecture}
For the loop $\mathrm{O}(1)$ model (see for example \cite{angel2021uniform} for a definition in a magnetic field) the problem may be more intricate since the lack of monotonicity appears stronger \cite{klausen2021monotonicity}. We note that in the limit $h \to \infty,$ the ghost edges are almost always open. In the limit, this changes the parity constraint of the marginal on the inner edges from percolation where all vertices are conditioned to have even degree into percolation where all vertices are conditioned to have odd degree. \\ 
\indent If we denote the critical $p$ for odd and even percolation by $p_{c,\text{odd}}$ and $p_{c, \text{even}}$ respectively, then if it is the case that  $p_{c,\text{odd}} >p_{c, \text{even}},$ this would be an illustration of how non-monotone the loop $\mathrm{O}(1)$ model is. 
We note that in the planar case $d=2$ and $h=0,$ it is proven \cite{GMM18} that $p_{c,even} =1-p_c(2,0)$, but to our knowledge, nothing is known about odd percolation. \\
\indent From this result it follows from the couplings (\cite[Exercise 36]{DC17}) that the single and double random currents also have the same phase transition for $d=2$. Using these increasing couplings, one can infer bounds on the Kert{\'{e}}sz line between the models. Thus, bounding the Kert{\'{e}}sz line for random currents or the  loop $\mathrm{O}(1)$ models may be a way to obtain better bounds on the  Kert{\'{e}}sz line for the random-cluster model than presented here and in \cite{Rui}. Studying the Kert{\'{e}}sz line for random currents may also shed some light upon the problem of whether the single current and double current have the same phase transition (\cite[Question 1]{DC16}).

\subsection*{Acknowledgements} 

 The first author acknowledges funding from Swiss SNF. The second author thanks the Villum Foundation for support through the QMATH center of Excellence (Grant No.10059) and the Villum Young Investigator (Grant No.25452) programs.  The authors would like to thank Ioan Manolescu for discussions and Bergfinnur Durhuus for comments on an early draft of the paper. Furthermore, the authors are very grateful to an anonymous referee for detailed comments.
 
\bibliographystyle{unsrt}

\bibliography{mono1bib}

\end{document}